\documentclass{article}
\usepackage{lineno,hyperref}
\usepackage[]{placeins}
\modulolinenumbers[5]
\usepackage{authblk}
\providecommand{\keywords}[1]{{\textit{Keywords---}} #1}
\usepackage{graphicx}
\usepackage[layoutwidth=7in, layoutheight=10in, layouthoffset=0.75in, layoutvoffset=0.5in, top=42pt,bottom=42pt,left=60pt,right=48pt, twoside, letterpaper]{geometry} %\usepackage[margin=1.5in]{geometry}
\usepackage{setspace}
\usepackage{color}
\usepackage{amsmath}
\usepackage{amssymb}
\usepackage{float}
\usepackage{mathtools}
\usepackage{multirow}

\title{Bayesian estimation of trend components within Markovian regime-switching models for wholesale electricity prices: an application to the South Australian wholesale electricity market}

\author[1,2]{Angus Lewis}%\thanks{\texttt{angus.lewis@adelaide.edu.au}}}
\author[1,2]{Nigel Bean}
\author[1,2]{Giang T.~Nguyen}

\affil[1]{The University of Adelaide, School of Mathematical Sciences}
\affil[2]{ Australian Research Council Centre of Excellence for Mathematical and Statistical Frontiers}

\begin{document}
\maketitle 
\begin{abstract}
We discuss and extend methods for estimating Markovian-Regime-Switching (MRS) and trend models for wholesale electricity prices. We argue the existing methods of trend estimation used in the electricity price modelling literature either require an ambiguous definition of an `extreme price', or lead to issues when implementing model selection \cite{janczura2013b}. The first main contribution of this paper is to design and infer a model which has a model-based definition of extreme prices \emph{and} permits the use of model selection criteria. 

Due to the complexity of the MRS models inference is not straightforward. In the existing literature an approximate EM algorithm is used \cite{janczura2012}. Another contribution of this paper is to implement exact inference in a Bayesian setting. This also allows the use of posterior predictive checks to assess model fit. We demonstrate the methodologies with South Australian electricity market. 
\end{abstract}

%\end{frontmatter}
 
%\tableofcontents 

\section{Introduction}
Electricity is a unique commodity due to the fact that it is not currently economically storable at scale, and demand is effectively inelastic. This causes electricity spot prices to exhibit interesting behaviours not seen in other markets, such as large price spikes and drops, negative prices, mean reversion, weekly, seasonal and yearly trends, and a strong dependence on weather and business activities \cite{becker2007,eydeland2003,gonzalez2005,janczura2013b,mount2006,seifert2007}. This makes modelling electricity prices a challenging problem.

The South Australian (SA) market is a particularly interesting case study due to its relative isolation, extreme hot weather and high penetration of non-baseload renewables -- approximately 49\% of SA's generation in 2018 was from non-baseload renewables, primarily wind \cite{aemo17report}.

A standard approach to modelling electricity prices decomposes the price series into a seasonal component, \(s_t,\) and a stochastic component, \(X_t\). The price at time \(t\) is then given by \(P_t=s_t+X_t\) or \(P_t = s_t X_t\). Furthermore, it is common to decompose \(X_t\) as a mixture of \emph{base} prices and \emph{price spikes}. In the existing literature, the seasonal and stochastic components are estimated separately. First, the seasonal component is estimated and removed from the price series, to create a stationary process, from which the stochastic component is then estimated \cite{benth2012,janczura2013b}. 

For wholesale electricity prices, estimating the seasonal component is made more complex by extreme price spikes which, due to their magnitude, will affect trend estimates \cite{janczura2013b}. One solution posed in the existing literature is to \emph{first} identify extreme observations and replace them with a ``normal'' value, \emph{then} estimate the seasonal component on this simplified data using standard techniques \cite{janczura2013b}. However, if the stochastic model is used to define extreme prices then some statistical model comparisons are not possible. 
To overcome this issue, we propose that the trend model is combined with the stochastic model, and that the two components are estimated jointly. This ensures that statistical model selection criteria are permissible and furthermore, can be used to compare both the stochastic and trend components.

We use cubic splines and wavelet-based models to capture long-term trends, and piecewise constant functions to capture weekly periodicities. The use of wavelet filtering for long-term trends, and piecewise constant models for weekly seasonalities, is common in the electricity pricing literature \cite{nowotarski2012,janczura2013b}, but cubic splines are less common. Since we combine the trend and stochastic models into one, as well as working in a Bayesian paradigm, we cannot apply wavelet filtering directly, instead we construct a wavelet-based regression model. This model is relatively complex due to the need for \emph{padding}, so we also introduce the simpler cubic spline-based model. 

To capture stochastic variations in electricity prices, we use a Markovian-Regime-Switching (MRS) model, which is an extension of the Hidden Markov Model (HMM) that allows for dependence between observations. For reviews on existing models for electricity prices see \cite{benth2008,eydeland2003,weron2014}. 
The MRS models used in the electricity pricing literature \cite{bierbrauer2007,eichler2013, janczura2010, noren2013,weron2009}, and considered in this paper, have a subtle but important difference in their specification compared to MRS models used in the wider literature: the MRS models in this paper have \emph{independent} regimes (see Section \ref{mrsmodel} for details). This specification is more suitable for electricity spot markets where prices return rapidly to the base level following a price spike. Independent regime models were introduced to electricity price modelling by Huisman and De Jong \cite{huisman2003}, and have since been popular \cite{bierbrauer2007,eichler2013, janczura2010, noren2013,weron2009}.

Inference for MRS models with independent regimes is not straightforward since the regime sequence is unobservable. In the electricity pricing literature, an approximation to the EM algorithm, which we will refer to as the \textit{EM-like} algorithm, is used for inference of these models (see Janczura and Weron \cite{janczura2012}, who extend work by Kim \cite{kim1994}). However, since this is an approximation to the EM algorithm, the theory of the EM algorithm is not valid and there is no promise of convergence, or error bounds, for this algorithm \cite{lewis2018}. 
We address this by working under a Bayesian paradigm and use augmented Markov Chain Monte Carlo (MCMC) methods for inference. Working in a Bayesian paradigm also alleviates identifiability issues related to estimation of shifted distributions compared to working in a maximum-likelihood context \cite{hill1963, lewis2018}. 

Once a model has been fitted, we must assess the assumptions underlying the model. Janczura and Weron \cite{janczura2013} suggest a method to assess goodness-of-fit when parameters are estimated using their EM-like algorithm. This is not directly applicable in our Bayesian setting. Instead, we adopt \emph{posterior predictive checks} \cite[Chapter 6]{gelman2004}. The logic behind posterior predictive checks is that if the model is good, then data replicated under the posterior predictive distribution should look similar to the observed data. By comparing the predictive distribution to observed data, we can assess if there are any significant deficiencies in the model. We also compute Bayes factors which we also use to compare models. 

To demonstrate our methodologies, we apply them to daily-average wholesale electricity prices in South Australia. Here we model daily average prices, as is common in the existing literature \cite{becker2007,erlwein2010,gonzalez2017,janczura2010,mount2006,regland2012,weron2009}.  

Throughout this paper we imprecisely use ``spot price'' to refer to the daily average spot price. We choose to model the raw daily-average price series, rather than log prices. There is some existing evidence to suggest that this is a reasonable approach \cite{weron2009}. Our justification is that to take the logarithm of prices, all prices need to be shifted upward so that they are positive. However, there is limited advice on how to shift the data, and the shift can have a large effect on the features of the price series. 

%From our modelling we find that a drop regime is unnecessary. We also suggest that two spike regimes should be included; one to capture typical spikes, and one to capture the very extreme spikes. Another observation is that there is a non-stationary variance effect; there appears to be a jump in volatility during 2016 that is sustained to the end of our dataset. 

%The structure of this paper is as follows. In Section~\ref{model} we introduce the seasonal component and the regime-switching model. Section~\ref{inference} presents our MCMC implementation, inference, and model checking methodologies. In Section~\ref{sadailyavemodelanalysis} we apply our models and methods to the South Australian electricity market, and in Section~\ref{conclusion} we make concluding remarks. The more technical details of the trend models and MCMC implementation are left to the Supplementary material.

\section{Model}\label{model}
Spot price evolution is a discrete-time process hence we use a discrete-time model. The model is composed of two parts, a seasonal component \(\{s_t\}_{t \in \mathbb N}\), capturing deterministic trends in electricity prices, and a stochastic component \(\{X_t\}_{t \in \mathbb N}\), capturing mean reversion, spikes and drops. We use an additive model; at time \(t\) the price, \(P_t\), is given by \(P_t = s_t + X_t\). 

Extreme prices in electricity markets can bias estimates of seasonal components if we do not treat them carefully. A solution proposed by Janczura \emph{et al.~}\cite{janczura2013b} is to use some method to identify and replace price spikes, and then estimate the trend model on the altered dataset. Once the trend model has been estimated it can be removed from the data, then the stochastic model estimated. For example, one method of spike identification used by Janczura \emph{et al.~}\cite{janczura2013b} is to regard any price further than \(3\) standard deviations above the mean of the data, as a spike. An alternative, also proposed by Janczura \emph{et al.~}\cite{janczura2013b}, is to iterate between estimating the seasonal component and estimating the stochastic component: a byproduct of estimating the stochastic component is a classification of observations into regimes, which can then be used to identify spikes. 

If a spike classification method which is independent of the stochastic model is used, such as classifying prices as spikes if they are three standard deviations or more away from the mean, then we can remove the trend from the data and estimate all stochastic components on the same detreded data set. Since the stochastic components all use the same detrended data, statistical (likelihood-based) metrics can be used to compare the stochastic component of the model. Issues with this method are that the choice of spike classification method is quite arbitrary, and we cannot use these metrics to compare trend components. 

Using the stochastic component to identify and replace spikes is attractive as it is \emph{model-based}, so there is an obvious definition of a price spike. Different specifications of the stochastic component will likely lead to different classifications of which prices are extreme, and therefore which prices must be treated before we estimate the trend component. Ultimately, the estimates of the trends will differ and therefore lead to different detrended datasets that we then use to estimate the stochastic components. The result is that the likelihoods of two different models are incomparable and therefore the use of statistical (likelihood-based) model comparisons such as the BIC or Bayes factors are invalid.

In this paper, we propose a different method. We combine the stochastic and trend models into one, and estimate them jointly. The advantages of this approach are that the classification of data as spikes is model-based, while also enabling statistical model comparison techniques for the stochastic components \emph{and} the trend components.

\subsection{The seasonal component, \(s_t\)}\label{season}
Electricity spot prices exhibit seasonality on weekly, seasonal and yearly scales. To capture this multi-scale seasonality, the seasonal component consists of two parts: a short-term component, \(g_t\), and a long-term component, \(h_t\). We present two different models for the long-term component, \emph{cubic splines}, and a \emph{wavelet-based} model. It is common to use \emph{wavelet filtering} to estimate long-term trends in the electricity pricing literature \cite{janczura2013b}, but cubic splines are less common. Some authors discount the use of splines due to the fact that they do not perform well for prediction, since it is not clear how to extrapolate spline models beyond the data \cite{nowotarski2012}. However wavelet filtering has similar issues. The appeal of cubic splines and wavelet models are that they are able to capture a diverse range of behaviours (unlike sinusoidal-based models, for example, which can capture periodic trends only). 

The general idea of wavelet filtering is to project a time series on to a set of functions which can represent low-frequency components of the data only. Specifically, for wavelet filtering the data is projected on to a set of \emph{scaling functions}. Due to properties of wavelet and scaling functions, the coefficients of the scaling functions can be computed as a \emph{linear filter} which is applied recursively to the data. The weights of the filter take values defined by wavelet and scaling basis functions. 

There are different possible choices of wavelet bases which give the filter different properties. Here, in line with existing literature, we choose to use Daubechies wavelets and apply the filter recursively 8 times \cite{weron2009}. In \cite{weron2009} Daubechies wavelets of order 24 are specified. However, we use Daubechies wavelets of order 8 due to computational demands of the algorithms used in this work. This serves to reduce the number of parameters of the wavelet component from 55 to 24, and also to reduce the amount of \emph{padding} required. In order to apply the wavelet filtering method, the data must be padded to make the data the correct length. Here we use \emph{symmetric} padding and extend the data at both ends by reflecting the signal (see the supplementary material for more details). 

Let the vector of data be \(\boldsymbol x = (x_1,...,x_T)'\), where \(\boldsymbol y'\) is the transpose of the vector \(\boldsymbol y\), and assume it is an appropriate length (or has been suitably padded). We show, in the supplementary material, that computation of the scaling function coefficients at level \(\widehat{\boldsymbol a}_j = (\widehat a_{1,j},...,\widehat a_{k,j})'\), can be represented as a matrix-vector product; 
\[\widehat{\boldsymbol a}_j = W_j \boldsymbol x,\] 
where \(W_j\) is a matrix of filter coefficients constructed in the supplementary material. We also show that, due to the properties of wavelet filtering, the coefficients \(\widehat{\boldsymbol a}_j\) are the least-squares estimates of the regression model \(\boldsymbol x = W_j' \boldsymbol a_j + \sigma \boldsymbol \varepsilon\), where \(\boldsymbol \varepsilon\) is a column vector of independent standard normal random variables. Motivated by this, we use the design matrix \(W_j'\) as a basis for our wavelet-based trend model. 

Similarly to how wavelet filtering is a projection on to scaling functions, we can also project the data on to a set of cubic spline functions. Let \(\eta_1<\eta_2<...<\eta_k\) be a set of points which are called \emph{knots}. Then cubic splines are piecewise cubic polynomials on the intervals \([\eta_1,\eta_2],[\eta_2,\eta_3],...,[\eta_{k-2},\eta_{k-1}],[\eta_{k-1},\eta_k]\), with continuous first and second derivatives at the boundaries of each interval, \(\eta_2,...,\eta_{k-1}\). In the supplementary material we show how to construct the design matrix of a cubic spline ordinary least squares regression model. We denote the design matrix of the cubic spline regression as \(C'\). Here we use \emph{B-splines} to construct the design matrix and place knots every 180 days (approximately two per year). An advantage of the cubic spline model is that no padding is required which simplifies the design matrix and computations. 

The other component of the trend model is the short-term component, which is made up of piecewise constant functions, and captures the average price fluctuations over the course of a week. That is, 
\begin{align*}g_t &= \beta_\text{Mon}\mathbb I (t \in \text{Mon}) + \beta_\text{Tue}\mathbb I (t \in \text{Tue}) + ...+ \beta_\text{Sun}\mathbb I (t \in \text{Sun}),\end{align*}
where \(\beta_\text{Mon},\beta_\text{Tue},...,\beta_\text{Sun},\) are the mean prices, after accounting for long-term trends, on Monday, Tuesday,..., Sunday, respectively and \(\mathbb I\) is the indicator. The design matrix for this component of the trend is denoted \(M'\) and defined in the supplementary materials. 

Putting this all together, the trend model has design matrix \(Z=\left[M'\mid W_j'\right]\) in the case of the wavelet-based model (which includes padding) and has design matrix \(Z=\left[M'\mid C'\right]\) in the case of the cubic spline based model. We denote the parameters of the trend model as \(\boldsymbol \gamma = (\gamma_1,...,\gamma_p)'\), therefore the trend is given by \(\boldsymbol s =  Z\boldsymbol \gamma\).

\subsection{Markovian-Regime-Switching models}\label{mrsmodel}
The Markovian-Regime-Switching (MRS) model is a generalisation of the hidden Markov model since MRS models allow for dependence between observations. MRS models are comprised of two components; a hidden \emph{regime} sequence, \(\{R_t\}_{t\in\mathbb N}\), and an \emph{observation} sequence, \(\{X_t\}_{t\in\mathbb N}\). The evolution of the regime process \( \{R_t\}_{t \in \mathbb{N} } \) is governed by a discrete-time Markov chain and is assumed to be unobservable. The regime process is completely defined by its state space, \(\mathcal{S} = \{1,2,...,N \}\), which corresponds to the set of regimes, a transition probability matrix, \(\boldsymbol{P}=(p_{ij}),\) \({i,j}\in \mathcal S\), and an initial distribution of the state of the chain. At each time \(t\), the observation, \(x_t\), follows a distribution that is known given the regime sequence up to time \(t\), \(\boldsymbol{R}_t = (R_1,...,R_t)'\), and all previous observations, \mbox{\((x_1,...,x_{t-1})'=:\boldsymbol{x}_{t-1}\)}. That is, the density function, \(f(x_t|\boldsymbol x_{t-1},\boldsymbol R_t)\), is known. Notice, unlike a hidden Markov model, the observations are \emph{not} conditionally independent given the hidden regimes. 

A key element of the independent regime MRS models used in the electricity price modelling literature is that observations generated by one regime are independent of observations generated from any other regime. That is, define \mbox{\(\mathcal A_i := \{X_t : R_t=i\}\)}, for \( i \in \mathcal S\), then \(\mathcal A_i\) and \(\mathcal A_j\) are independent for \(i\neq j\). One way to think about the evolution of these models is as follows. Consider \(M\) random processes evolving simultaneously, each of which correspond to a regime in the MRS model, and each of which produces a realisation at all time points. At each time \(t\) we observe one of these processes only, and which process is observed is determined by the hidden regime process, \(\{R_t\}\). That is, at time \(t\), we observe the \(i^\text{th}\) process if \(R_t = i\), and all other realisations are unobserved. Due to this assumption, the distributions \(f(x_t|\boldsymbol x_{t-1},\boldsymbol R_t)\) may depend on the history of \(\boldsymbol R_t\). 

\subsection{The full model} 
We wish to combine the MRS and trend models into one so that we may estimate them jointly. We specify two types of regimes, \emph{base} regimes, denoted as \(\{B_t^{(i)}\}_{t\in \mathbb N}\), \mbox{\(i= 1,...,n\),} which are AR(1) processes and include trend components, and non-base regimes, \(\{Y_t^{(i)}\}_{t\in \mathbb N}\), \mbox{\(i=n+1,...,N\),} which are i.i.d~processes and do not have trend components. The processes, \(\{B_t^{(i)}\}_{t \in \mathbb{N}}\) have the form
\begin{align}\label{ar1eqn} B_t^{(i)} &= \phi_i (B_{t-1}^{(i)}-s_{t-1}) + s_t + \sigma_i \varepsilon_t^{(i)} \\&=\phi_i (B_{t-1}^{(i)}-\boldsymbol z _{t-1}\boldsymbol \gamma) + \boldsymbol z _{t}\boldsymbol \gamma+ \sigma_i \varepsilon_t^{(i)} \nonumber ,\end{align}
where \(\boldsymbol \gamma, \phi_i\) and \(\sigma_i\) are parameters and \(\{\varepsilon_t^{(i)}\}_{t \in \mathbb{N}}\) is a sequence of i.i.d.~\(N(0,1)\) random variables, \(i=1,2,...,n\). The terms \(\boldsymbol z_{t-1}\) and \(\boldsymbol z_{t}\) are the rows of \(Z\) corresponding to times \(t-1\) and \(t\), respectively. When multiple AR(1) regimes are included in a model we restrict \(\sigma_i<\sigma_{i+1},\) for all \(i = 1,2,...,n\). This condition serves to identify the model. Rearranging Equation~(\ref{ar1eqn}) to \[B_t^{(i)}-s_t  = \phi_i (B_{t-1}^{(i)}-s_{t-1}) + \sigma_i \varepsilon_t^{(i)}\]
shows that this model corresponds to modelling the noise around the trend as a stationary AR(1) process. 

For all other regimes, let \(\{Y_t^{(i)}\}_{t \in \mathbb N}\) be a sequence of i.i.d.~random variables for each \(i=n+1,...,N\). These regimes are labelled as either \emph{spikes} or \emph{drops} depending on their nature. 

The largest models that we consider have \(N = 5\) regimes: two base regimes, two spike regimes, and a drop regime. Other models considered are subsets of this largest model. The inclusion of two base regimes in our modelling is motivated by our observations of a structural change in volatility in 2016 in the dataset. As for spike regimes, one spike regime captures typical spikes and another captures extraordinary spikes, which is motivated by our observation that a single spike regime is unable to capture extreme price spikes in the SA market data. The drop regime captures large downward price movements which can occur when the market is oversupplied. 
Therefore, the largest model that is considered takes the following form, 
\begin{equation*}
P_t = \begin{cases} 
B_t^{(1)} & \text{if } R_t = 1, \\
B_t^{(2)}& \text{if } R_t = 2,\\
Y_t^{\left(3\right)} & \text{if } R_t = 3, \\
Y_t^{\left(4\right)} & \text{if } R_t = 4, \\
Y_t^{\left(5\right)} & \text{if } R_t = 5,
\end{cases}
\end{equation*}
where \(P_t\) is the price at time \(t\), \(B_t^{(i)}\), \(i=1,2,\) are as in Equation~(\ref{ar1eqn}), and \(Y_t^{\left(j\right)},\) \(j=3,4,5,\) are i.i.d.~random variables. We consider two different specifications for Regimes 3 and 4 (spike regimes): a shifted log-normal and a shifted gamma distribution,
\begin{align*}Y_t^{\left(i\right)}-q_i \sim F_i\left(\mu_i,\sigma_i^2\right)\text{ for } i=3,4,\end{align*}  
where \(F_i\) is either a log-normal or a gamma distribution, and \(q_i\) is a shifting parameter. Note that we parameterise the Gamma distribution with shape parameters \(\mu_i\), and scale parameter \(\sigma_i^2\). Regime~5, the drop regime, follows an i.i.d.~shifted log-normal distribution but with the direction of the distribution reversed; 
\begin{align*}q_5-Y_t^{\left(5\right)} \sim LN\left(\mu_5,\sigma_5^2\right),\end{align*}
and is included to capture large downward price movements.
The shifting parameters \(q_3,q_4\) and \(q_5\) can be prespecified \cite{janczura2012}, although we leave them to be inferred. We restrict the support of their prior distributions for identifiability and interpretability. 

\begin{figure}
	\centering
	\includegraphics[width = 0.6\textwidth, trim = {20 0 20 26.5}, clip]{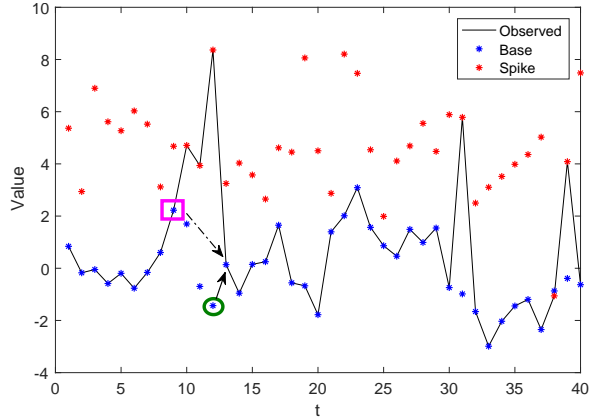}
	\caption{Simulation of an MRS model with two independent regimes; an AR(1) base regime and an i.i.d~spike regime. To write down the distribution of the observation at time \(t=13\), we would like to know either the base regime value immediately before it, circled in green, or the time that the last observed price from that regime occurred, highlighted by the pink box. }
		\label{sim}
\end{figure}

As mentioned earlier, the regimes of the MRS model are assumed to be mutually independent. This feature allows for the true spike-like behaviour observed in electricity markets, because it ensures that the return to base prices after spikes is immediate. 
However, this assumption does create some complicated dependence between observations of the AR(1) regimes. To determine the distribution of the AR(1) processes at time \(t\), \(B_t^{(i)}\), we either need to know the previous price from that regime, \(B_{t-1}^{(i)}\) (which may or may not have been observed), or, the time at which the last observation from that regime occurred, in which case we can integrate out the missing values of the AR(1) process between \(B_{t}^{(i)}\) and the last observed price from that regime, to construct its distribution. We take the latter approach. To illustrate, Figure \ref{sim} shows a simulation of a two-regime model with one AR(1) regime and one i.i.d.~spike regime.  

\subsection{Prior distributions}
To complete our model we need to specify prior distributions for the parameters of our model. We look, where possible, to specify uninformative priors. However, for some parameters we use the prior distributions to enforce certain behaviours. 

We use the prior distribution to restrict the support of the parameters \(q_i\) of the shifted spike and drop distributions. If left unrestricted, then these parameters can make the spike and drop regimes capture base prices, rather than the behaviour they were designed to capture. For example, for spike distributions the shifting parameter may become negative and the corresponding regime captures prices that we would not logically classify as spikes. Janczura and Weron \cite{janczura2009} suggest setting \(q_5\), the shifting parameter of the drop regime to the \(25^\text{th}\) percentile of the data and \(q_3\), the shifting parameter of the spike regime to the \(75^\text{th}\) percentile of the data if they cannot be estimated. Motivated by this, we place a uniform prior on these parameters around these values. Specifically \(q_5\sim U[Q_{0.01},Q_{0.33}]\) and \(q_3\sim U[Q_{0.66},Q_{0.99}]\), where \(Q_{\alpha}\) is the \(\alpha\times 100\%\)th percentile of the data. When two spike regimes are specified we set the prior of the shifting parameter of the second spike distribution to \(q_4\sim U[Q_{0.9},Q_{0.99}]\).

Estimating shifted log-normal and shifted gamma distributions is not straightforward as noted by Hill \cite{hill1963} and Johnson and Kotz \cite{johnson1994}. This is due to the fact that there are points where the likelihood may take arbitrarily large values. For example, if the shifting parameter is equal to a data point, \(q_i=x_t\), then the likelihood of the shifted log-normal distribution can be arbitrarily large if the variance tends to 0. 

For the shifted log-normal distribution the variance parameters, \(\sigma_i^2\), have a prior distribution with densities \(f(\sigma_i^2)\propto \frac{1}{\sigma_i^2}\), \(i=3,4,5\), and support \(\sigma_i^2 \in [0.1,10\times s^2]\), where \(s\) is the standard deviation of the data. A key feature here is that we restrict \(\sigma_i^2\) to not be arbitrarily small. This form of prior specification is often justified by the fact that, if the support of this prior were \((0,\infty)\), then this would be the Jefferies prior for this distribution, which is invariant to reparameterisation, and also \(\log\sigma^2\) has a uniform distribution on \(\mathbb R\). The restriction of the prior to a finite interval is so that the prior can be normalised and so that \(\sigma_i^2\) cannot become arbitrarily small preventing the likelihood from taking arbitrarily large values. We set the prior distribution for the location parameter of the shifted log-normal to a normal distribution \(N(0,10\times s)\) where \(s\) is the standard deviation of the data. 

It can also be shown that the likelihood of the shifted gamma distribution can be infinite when \(q_i=x_t\) and the shape parameter \(\mu_i\leq1\). In addition, the density of the gamma distribution is equal to 0 at \(x=q_i\) only if \(\mu_i>1\). Therefore, when included in our MRS models, if we allow \(\mu_i\leq 1\), the shifted gamma spike regimes will have a discontinuous peak at \(x=q_i\), and most of their mass will lie near this point. Since we do not expect price spikes to be largely clumped around the shifting parameters \(q_i\), we force the shifted gamma distributions to have a peak away from the boundary using the prior distribution of \(\mu_i\). 

For the shifted gamma distributions, the prior distributions for the parameters \(\mu_i\) are shifted inverse-gamma distributions, which are chosen so that we can enforce desirable behaviour on the distribution in these regimes. These prior distributions are shifted to have support from \(1\) to \(\infty\) (rather than 0 to \(\infty\)) to ensure that the shape parameters \(\mu_i\) are strictly greater than 1. Furthermore, the inverse-gamma prior distributions are given shape parameter \(\alpha = 3 \) (the smallest integer such that the prior distribution has finite variance) and scale parameter \(\beta = 5.5\), so that the prior distribution has its mode at \(2.5\). This follows advice from Johnson and Kotz \cite{johnson1994} who, in a maximum-likelihood context, observed that issues related to the likelihood becoming large for \(\mu_i<1\) arise when \(\mu_i\) is near 1 and they advise against maximum likelihood estimation of the shifting parameter \(q_i\) when \(\mu_i< 2.5\). The issues discussed by Johnson and Kotz \cite{johnson1994} may still occur in a Bayesian setting, but may be less prominent due to the prior distribution \cite{hill1963, lewis2018}. For the scale parameter of the shifted gamma distribution we use the same form of prior distribution as for those used for variance terms throughout; \(f(\sigma_i^2)\propto \frac{1}{\sigma_i^2}\) with support \([0.1,10\times s^2]\), where \(s\) is the standard deviation of the data. 

For the transition probabilities of the hidden Markov chain we assign a Dirichlet prior distribution with all parameters equal to 1: \[(p_{1j},\dots,p_{Nj})  \sim Dirichlet(1,\dots,1).\]
This is actually a uniform prior distribution subject to a unit sum condition

The prior distribution of the correlation parameters \(\phi_i\) is \(U(-1,1)\), uniform on the interval \((-1,1)\). This captures the fact that \(|\phi_i|<1\) is a requirement for an AR(1) process to be stationary. 

The variance terms of the base regimes have a distribution \(f(\sigma_i^2)\propto \frac{1}{\sigma_i^2}\) \(i=1,2\), with support \([1,10\times s]\) where \(s\) is the standard deviation of the data. The justification for this is the same as for the prior distributions of the log-normal distributions specified above. 

For the parameters of the trend models, \(\boldsymbol \gamma\), we assign independent, normally distributed prior distributions with mean 0 and standard deviation \(10\times s^2\).

As a small investigation on the sensitivity of our modelling to the choice of priors, we replaced the prior distributions of the form \(f(\sigma_i^2)\propto \frac{1}{\sigma_i^2}\) with uniform prior distributions with the same support and observed no substantial change in our conclusions.  

\section{Inference}\label{inference}

\subsection{Bayesian inference for the MRS model using MCMC} \label{mcmc}
To estimate the parameters of our MRS models we use an adaptive, data-augmented, block Metropolis-Hastings algorithm to sample from the posterior distribution. In the following we use the notation for a parameter vector, \(\boldsymbol\theta \in \Theta\), where \(\Theta\) is the parameter space, the prior distribution, \(\pi(\boldsymbol\theta)\), the posterior distribution, \(p(\boldsymbol\theta | \boldsymbol x)\), the likelihood, \(p(\boldsymbol{x}|\boldsymbol\theta)\) and define \(p(\boldsymbol{x}) = \int_\Theta p(\boldsymbol{x}|\boldsymbol\theta)\pi(\boldsymbol\theta)d\boldsymbol \theta\).

\paragraph{Likelihood} Due to the specification of the MRS model, the distribution of observations is determined by knowledge of the hidden regime sequence and previous price data, therefore we must write the likelihood as a marginal distribution. Let \(\mathcal S=\{1,...,N\}\) be the state space of the regime process and \(\boldsymbol x = (x_1,...,x_T)'\) be our sequence of observed prices. Then the likelihood can be written as
 \begin{equation}\label{likelihoodeq} p(\boldsymbol x|\boldsymbol\theta) = \sum_{\boldsymbol R \in \mathcal S^T} p(\boldsymbol x , \boldsymbol R | \boldsymbol\theta) = \sum_{\boldsymbol R \in \mathcal S^T} p(\boldsymbol x | \boldsymbol\theta, \boldsymbol R)p(\boldsymbol R | \boldsymbol\theta).\end{equation}
The number of sequences in \(\mathcal S^T \) is \(N^T\). For most realistic datasets it is impossible to enumerate all \(N^T\) possible regime sequences so we cannot naively compute the sum on the right-hand side of Equation (\ref{likelihoodeq}).

Recall Figure \ref{sim} and the related discussion about the complex dependence structure between observations. Due to this dependence structure the standard EM algorithm for MRS models is not valid (see \cite{hamilton1989,hamilton1990} for the algorithm and \cite{janczura2012,lewis2018} for a discussion of why it is not applicable to this problem). In the existing literature an \textit{EM-like} algorithm is used for inference \cite{janczura2012} of these electricity price models. However, the EM-like algorithm is an approximation to the EM algorithm, hence the theory of the EM algorithm is not valid and there are no known error bounds, or promise of convergence \cite{lewis2018}. Janczura and Weron \cite{janczura2012} provide some simulation evidence to show that their algorithm works well, but it is not too hard to find examples where it fails \cite{lewis2018}. 

To resolve the inference problem we work in a Bayesian setting and use data-augmented MCMC methods. Thus, our methodology is backed by a vast literature which suggest we will achieve reasonable parameter estimate. The data-augmentation allows us to work with \(p(\boldsymbol x | \boldsymbol\theta,\boldsymbol R) \) instead of the likelihood, overcoming the computation issue in Equation (\ref{likelihoodeq}). 

Let \(s^t_i \) be the length of time since the regime was last in state \(i\) before time \(t\), and the regime at time \(t\) is \(i\). For example, the event \(\{s^t_{1}=k\} = \{k\in\mathbb N\mid R_t = 1, R_{t-1} \neq 1, ...,R_{t-k+1} \neq 1, R_{t-k} = 1\} \). Then we can write \(p(\boldsymbol x| \boldsymbol\theta,\boldsymbol R)\) as
\begin{align*} 
p(\boldsymbol x |\boldsymbol\theta,\boldsymbol R)&=  \prod_{i=1}^2  p_{B^{(i)}}(x_1|\boldsymbol\theta)^{\mathbb I (R_1 = i)} 
\prod_{i=1}^2 \prod_{t = 2}^{T} \prod_{k = 1}^{t-1} p_{B^{(i)},k}(x_t|x_{t-k},\boldsymbol\theta)^{\mathbb I (s_{i}^t = k)}
\\&{}\quad\times \prod_{i=3}^5\prod_{t = 1}^T p_{Y^{(i)}} (x_t|\boldsymbol\theta)^{\mathbb I (R_t = i)}.
\end{align*}
Here \(p_{B^{(i)},k}(\cdot|x_{t-k},\boldsymbol\theta)\) is the density of the \(i\)th AR(1) process, given that the last observation from the \(i\)th process, was \(x_{t-k}\) and was \(k\) lags ago, \(p_{Y^{(i)}}(\cdot|\boldsymbol\theta)\) is the density when in the \(i\)th i.i.d.~regime. For simplicity we assume that \(R_1 = 1\), and, as there is no prior price information, \(p_{B^{(i)}}(x_k|\boldsymbol\theta) = 1\) for \(i=1,2\), where \(k\) is the first time in regime \(i\). That is, we effectively ignore the first observation from the autoregressive regimes. For large sample sizes, this assumption is reasonable and has minimal effect. 

The decision to evaluate \(p(\boldsymbol x |\boldsymbol\theta,\boldsymbol R)\) instead of the likelihood means that we infer the posterior distribution of the parameters \emph{and} the unobserved regime sequence. That is, we infer
\begin{align*}p( \boldsymbol\theta,\boldsymbol R|\boldsymbol x) = \frac{p(\boldsymbol x | \boldsymbol\theta,\boldsymbol R)p(\boldsymbol\theta, \boldsymbol R)}{p(\boldsymbol{x})} \propto p(\boldsymbol x |  \boldsymbol\theta,\boldsymbol R)p(\boldsymbol R|\boldsymbol\theta)\pi(\boldsymbol\theta).\end{align*}
The marginal distribution \(p(\boldsymbol\theta | \boldsymbol x)\) that we seek is extracted by integrating the posterior \(p(\boldsymbol\theta, \boldsymbol R|\boldsymbol x)\) over \(\boldsymbol R\). Conveniently, when using the Markov chain Monte Carlo (MCMC) methods that we use here, this integration is equivalent to simply ignoring the sampling of \(\boldsymbol R\) from our MCMC output. Thus, data augmentation enables us to construct an MCMC algorithm that efficiently samples from the posterior distribution but at the cost of also requiring us to infer the sequence of regimes, \(\boldsymbol R\).

\subsection{Model checking} \label{check}
We use \emph{posterior predictive checks} (PPCs) for model checking and comparison. PPCs are typically used as a flexible tool for model checking in Bayesian settings \cite[Chapter~6]{gelman2004}. The general idea is to sample parameters from the posterior distribution and then use these to calculate statistics from the observed data. This is repeated for many samples from the posterior and we look to see if, overall, our model looks reasonable. 

More specifically, for the implementation of PPCs in this context, we sample a parameter vector \(\boldsymbol\theta^*\) and a regime sequence \(\boldsymbol R^* \) from the posterior distribution. Sampling the hidden \(\boldsymbol R^*\) allows us to classify observations into regimes, and then calculate statistics for each regime individually. To test distributional assumptions we calculate the standardised residuals for each regime, which is straightforward for the i.i.d.~regimes but requires a little more thought for the base regimes. The regime sequence \(\boldsymbol R^*\) is used to classify points into base regime~\(i\), which gives a partially observed AR(1) process. For each point classified into base regime~\(i\), except the first point, the standardised residual, \(r_t^{(i)}\), is constructed using 
\begin{align}\label{residenq}r_{t}^{(i)} = \frac{x_t-\boldsymbol z_t \boldsymbol \gamma^* - (\phi_i^*)^{s_i^t}\left(   x_{t-{s_i^t}}-\boldsymbol z_{t-s_i^t} \boldsymbol \gamma^* \right)}{\sigma_i^*\sqrt{\cfrac{1-(\phi_i^*)^{2s_i^t}}{1-\left(\phi_i^*\right)^2}} },\end{align}
where \({s_i^t}\) is number of lags since the process was last in regime \(i\) before time \(t\), and is determined by \(\boldsymbol R^*\).

For each regime the residuals are used to perform the following model checks:
\begin{enumerate}
\item Construct QQ plots and compare to distributional assumptions.
\item Plot residuals against \(t\), to check for constant variance across time.
\item Plot residuals against the absolute value of the last observed value in that regime, i.e.~for the base regime~\(i\), plot \(r^{(i)}_{t}\) against \(|x_{t-{s_i^t}}|\) where \(\boldsymbol R^*\) determines \({s_i^t}\). We use this plot to check that the variance of the process does not depend on the magnitude of prices. 
\end{enumerate}
In Section~\ref{sadailyavemodelanalysis}, for the sake of brevity, we only provide a single representative plot from our PPCs. 

PPCs are easy to implement but, as with any checking procedure, we should be aware of their interpretation and drawbacks. A word of caution is in order. PPCs are useful to determine if our model fails in any obvious way but, due to the fact that we use the same data to fit and check the model, these procedures tend to make models look better than they actually are. 

As another model check we also plot the price series and a classification of observed prices into a regime estimated under the model. This allows us to visually assess which regime captures which data points to see if the model is behaving as we expect.

We also compute Bayes factors to compare models. For two models \(M_1\) and \(M_2\), the Bayes factor to compare these models is the ratio \(K=\cfrac{p(\boldsymbol x\mid M_1)}{p(\boldsymbol x\mid M_2)}\). That is, the ratio of the probability of the observed data under \(M_1\) compared to the probability of the data under \(M_2\). Thus, if \(K\) is greater than 1, then the probability of the observed data is greater under model \(M_1\) than under \(M_2\), thus \(M_1\) would be preferable. The quantity \(p(\boldsymbol x\mid M_1)\) is approximated within the implementation of the MCMC procedure. 
\FloatBarrier

\section{Application to the South Australian electricity market}\label{sadailyavemodelanalysis}
The National Electricity Market (NEM) is comprised of five interconnected states of Australia that also act as price regions: Queensland, New South Wales (including the Australian Capital Territory), South Australia, Victoria, and Tasmania. Each state has its own generation capacity and can also import/export electricity via interconnectors between states. \emph{Dispatch} prices for each region are set every 5 minutes in an auction process managed by the Australian Energy Market Operator (AEMO), and the \emph{spot price} is the average over 30 minutes of the dispatch prices, resulting in 48 realisations of the spot price process per day. The spot price is the price at which transactions are settled. 

\begin{figure}
\begin{center}
\includegraphics[width = 1\textwidth, trim={100 0 100 0}]{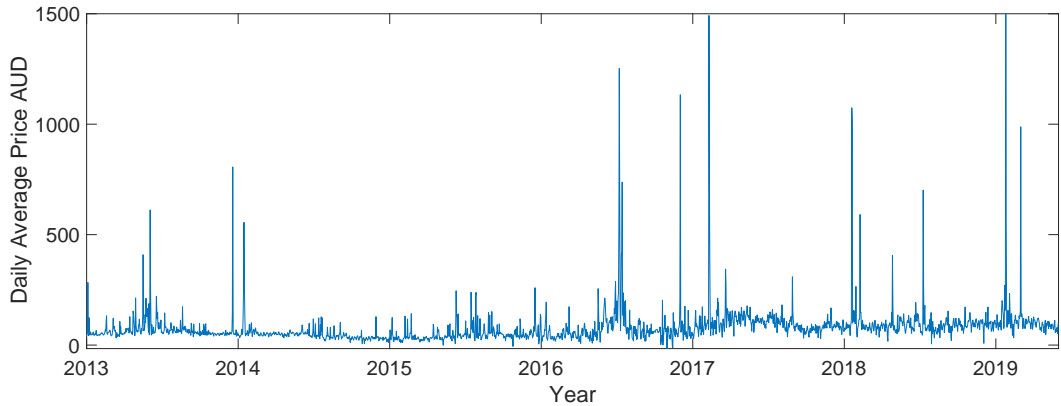}
\caption{The daily average wholesale electricity spot price for South Australia for the period from the \(1\)st of January 2013 to the \(31\)st of May 2019, quoted in \$AUD per Megawatt hour. The plot has been truncated at \$1500 so that details are clearer. The price which has been cut off in the truncated plot is a price of \$3359.81 which occurred in 2019.}
\label{dailyavedata}
\end{center}
\end{figure}

The South Australian electricity market is a particularly interesting case study due to a number of factors including its relative isolation, occasional extremely hot weather, and generation mix -- in 2016 SA had 39.2\% of is total generation come from wind farms, 50.5\% from gas and 9.2\% residential solar panels \cite{aemo17report}. 
Our dataset consists of 112416 half hourly electricity spot prices from the South Australian electricity market (available at the AEMO website \cite{aemo}) for the period  00:00 hours, \(1\)st of January 2013 to 23:30 hours \(1\)st of May 2019. This gives us a dataset of 2342 daily average price observations to which we fit our model. To be clear, we calculate the daily average prices as the arithmetic mean of the 48 spot prices in a day. The data that we model is plotted in Figure~\ref{dailyavedata}.

We fit the models discussed above to the South Australia dataset, using our MCMC algorithm to sample from the posterior distribution. For each model, 4 MCMC chains of length 2,000,000 were generated and the first 500,000 samples of each chain were used for adaption and discarded as burn-in.\footnote{Code available online at \url{https://github.com/angus-lewis/MRSMCMC}} First, we discuss the two trend models introduced above, and then the MRS models.

\paragraph{Trend model}
Recall the two trend models introduced earlier in the paper; the wavelet-based model, and the cubic spline-based model. It turns out that, under the model checking procedure that we follow below, the choice of trend model does not significantly affect the conclusions about the distributional assumptions of the model. That is, the distributional assumptions appear to be sufficient or deficient in the same ways for either trend model. Therefore to simplify this discussion we will focus on the trend model for log-normally distributed spikes only.

\begin{figure}
\begin{center}
\includegraphics[width = 1\textwidth,trim = {90 20 100 40}, clip]{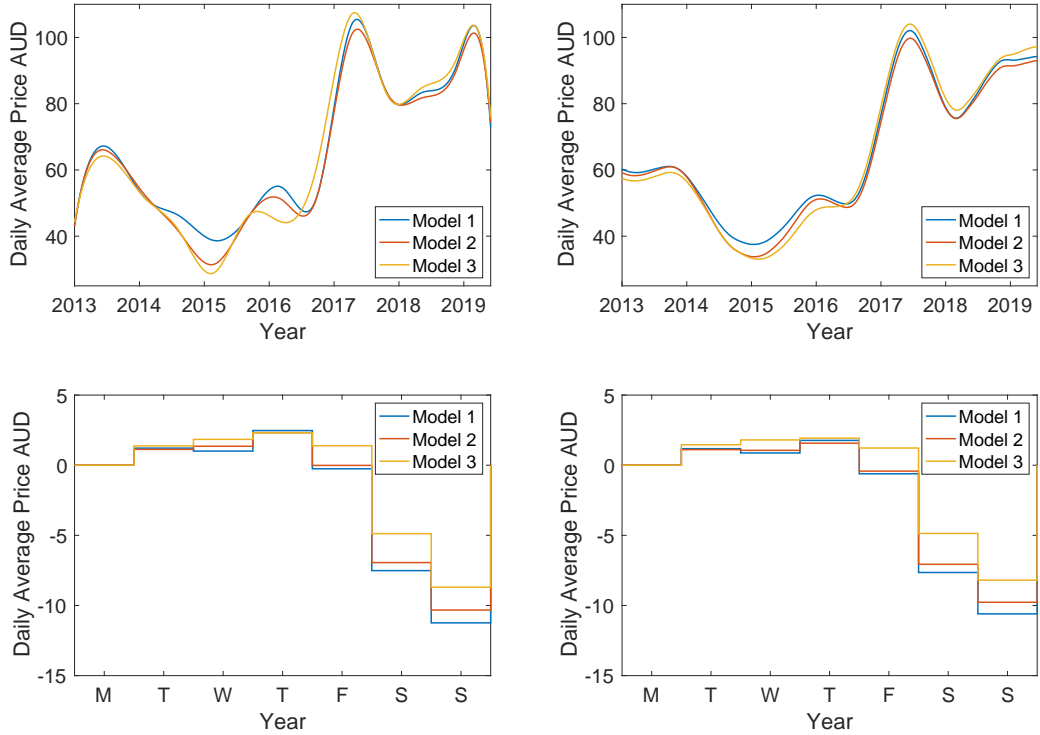}
\caption{Wavelet-based (left) and spline-based (right) posterior mean long-term (top) and short-term (bottom) seasonal components for stochastic Models 1-3. The short-term trend models have been adjusted so that they represent the average price on a given day of the week compared to the average price on a Monday.}
\label{fig:trends}
\end{center}
\end{figure}

In Figure \ref{fig:trends}, posterior mean estimates of the short-term and long-term trends are shown for stochastic Models 1-3 (defined below) for both the spline and wavelet models. The difference between the estimated mean long-term trend components is most notable at the edges of the dataset. This is not unexpected, since the wavelet-based model incorporates edges effects due to padding, while this is not an issue for the spline-based model. Both trend models, however, seem to capture the main features of the data, a trough in 2015 and a peak in 2017. 

Notable, Model 1 suggests that the trough in prices in 2015 was not as low as the other models suggest. However, this can be explained by the fact that Model 1 includes a drop regime to capture sudden price drops. However, for this model, rather than capture sudden price drops, the drop regime captures the period of low prices in 2015. In contrast Models 2 and 3 do not include drop regimes and so we see a lower trough in 2015 for these models. 

There is not much difference between estimates of the short-term trend component for the wavelet and spline based models. 

We can use Bayes factors to compare the cubic-spline based and wavelet based models. Table \ref{table: bayes factors trends} contains the natural logarithm of the Bayes factor comparing the the wavelet-based model to the cubic-spline based model for stochastic models 1-3.
\begin{table}[h!]
\caption{The natural log of the Bayes factor comparing the wavelet-based and cubic-spline based trend models for stochastic models 1-3. }
\centering
\begin{tabular}{l|cc}
&Wavelet vs.~Cubic-spline\\\hline
Stochastic model 1 & 6.59\\
Stochastic model 2 & 12.1\\ 
Stochastic model 3 & 34.1
 \end{tabular}\label{table: bayes factors trends}
\end{table}Table \ref{table: bayes factors trends} shows that the cubic-spline model is favourable in every case. Furthermore, the cubic-spline based model is simpler to implement and interpret hence we recommend the cubic spline based model is used. The results presented henceforth are all based on the spline-based model.

\paragraph{MRS model}To estimate the MRS elements of the model we take a stepwise approach to model building. Starting with a 3-regime model we fit it to the data and then interrogate the model using PPCs to find which aspects of the data the model is failing to capture. Then we add or remove elements of the model to attempt to improve model fit. This is repeated until a suitable model is found.

\textbf{Model 1.} First consider the following 3-regime model.
\begin{equation*}\label{eqnM1}
X_t = \begin{cases} 
B_t^{(1)} & \text{if } R_t = 1, \\
Y_t^{\left(3\right)} & \text{if } R_t = 3, \\
Y_t^{\left(5\right)} & \text{if } R_t = 5, \tag{Model 1}
\end{cases}
\end{equation*}
which is a reduced model of that introduced in Section~2.2. Regime \(Y_t^{(3)}\) follows a shifted log-normal distribution and captures price spikes, and \(Y_t^{(5)}\) follows a shifted and reversed log-normal distribution, to capture price drops. 

\begin{figure}
\begin{center}
\includegraphics[width = 1\textwidth]{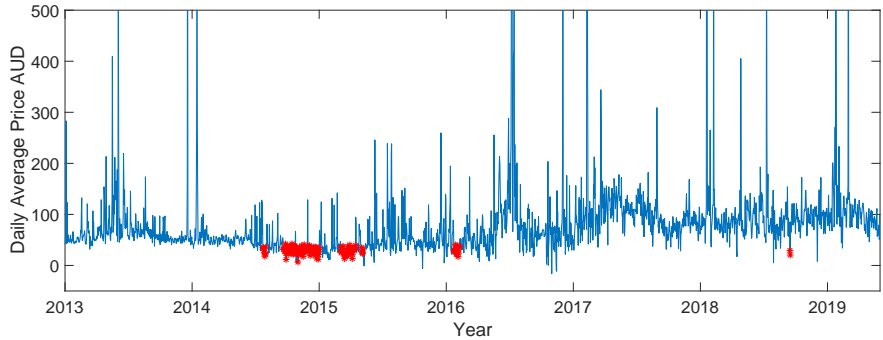}
\caption{The daily average wholesale electricity spot price for South Australia for the period from the \(1\)st of January 2013 to the \(31\)st of May 2019 quoted in \$AUD per Megawatt hour. The points highlighted in red were classified in to the drop regime, Regime 5, of \ref{eqnM1}, using the following classification rule. If an observation at time \(t\) has the greatest posterior probability of being in the drop regime, that is \(\mbox{arg}\max_j P(R_t=j|\boldsymbol x)=5\), then the observation is classified as coming from the drop regime. Clearly the drop regime is not capturing the behaviour it was designed to capture. }
\label{fig: classificationM1Drop}
\end{center}
\end{figure}

\begin{figure}
\begin{center}
\includegraphics[width = 0.5\textwidth, trim = {0 0 0 22}, clip]{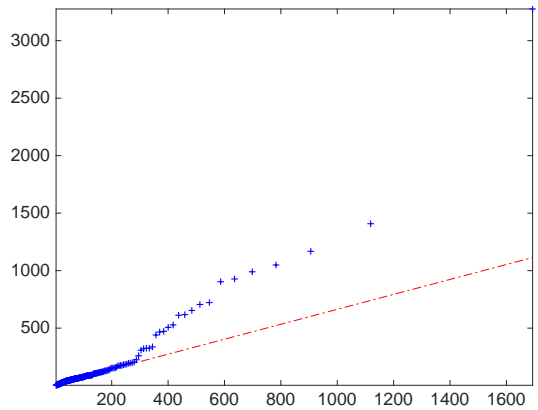}
\caption{A representative posterior predictive check QQ plot of the spike process, Regime 3, in \ref{eqnM1}. Due to the curvature in the points on the plot, the assumption that spikes follow a log-normal distribution is not valid.  }
\label{fig: LNQQplotM1}
\end{center}
\end{figure}

Fitting this model to the data we notice two aspects where the model clearly fails to model the data appropriately. The drop regime, Regime 5, does not capture large downward price movements, as expected. Rather, it captures a period of low `base' prices around the end of 2014 and in to the first half of 2015, as shown in Figure \ref{fig: classificationM1Drop}. Also, the assumption that the spikes follow a shifted log-normal distribution appears to be violated, which is indicated by the posterior predictive checks in Figure \ref{fig: LNQQplotM1}. Here, Figure \ref{fig: LNQQplotM1} includes just one representative posterior predictive check plot only, for brevity. 

For these reasons a second spike regime is added to the model -- one with a larger shifting parameter to capture extreme spikes -- while also removing the drop regime.

\textbf{Model 2.}
This model therefore includes one base regime,~\(B_t^{(1)},\) two spike regimes, one for typical spikes, \(Y_t^{(3)}\), and another for extreme spikes, \(Y_t^{(4)}\), and no drop regime:
\begin{equation*}\label{eqnM2}
X_t = \begin{cases} 
B_t^{(1)} & \text{if } R_t = 1, \\
Y_t^{\left(3\right)} & \text{if } R_t = 3, \\
Y_t^{\left(4\right)} & \text{if } R_t = 4, \tag{Model 2}
\end{cases}
\end{equation*}
where \(Y_t^{(3)}\) and \(Y_t^{(4)}\) follow shifted log-normal distributions.

\begin{figure*}
	\centering
	\includegraphics[width = 0.475\textwidth, trim = {0 0 0 22}, clip]{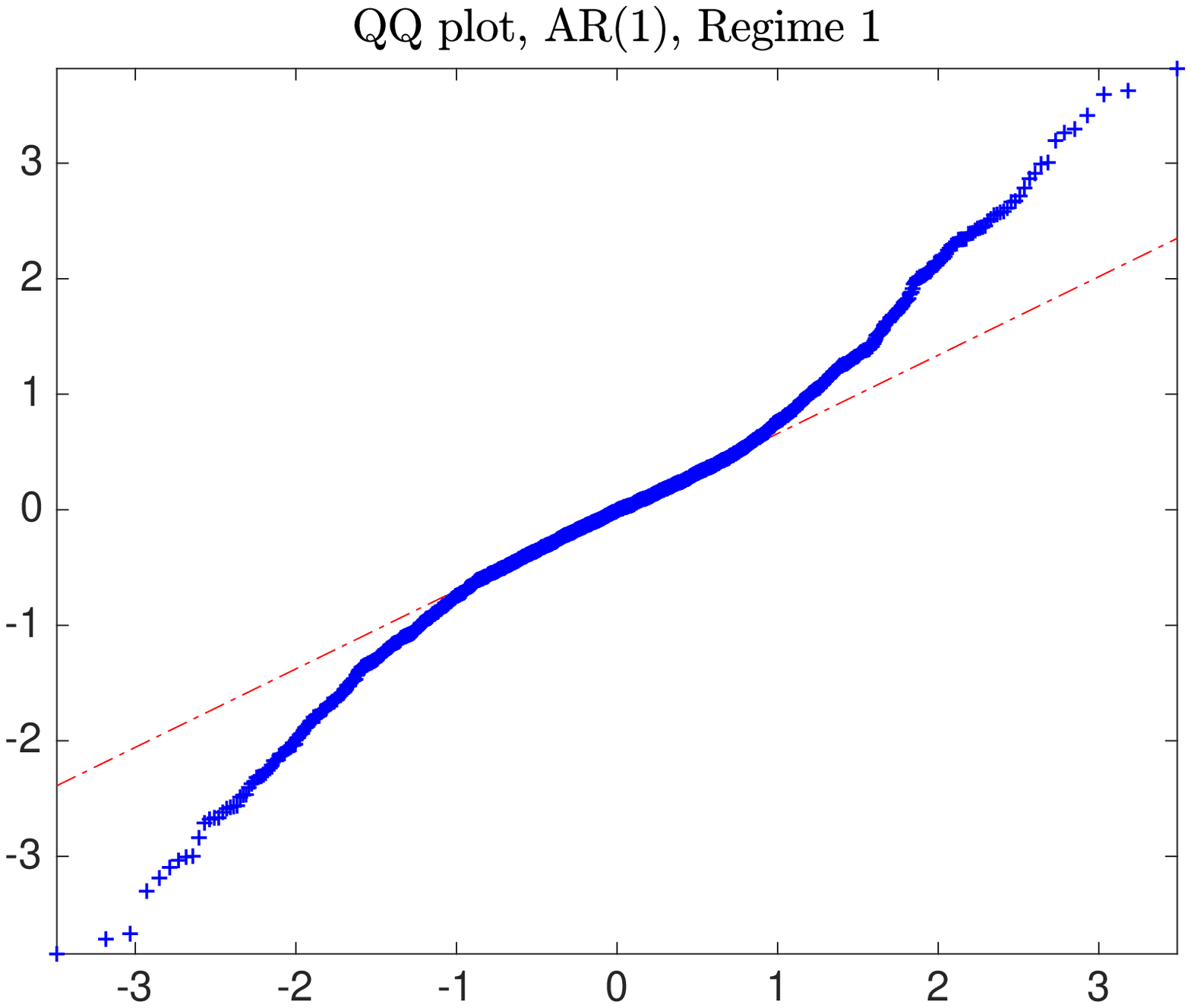}\hspace{4mm}\includegraphics[width = 0.49\textwidth, trim = {0 0 0 22}, clip]{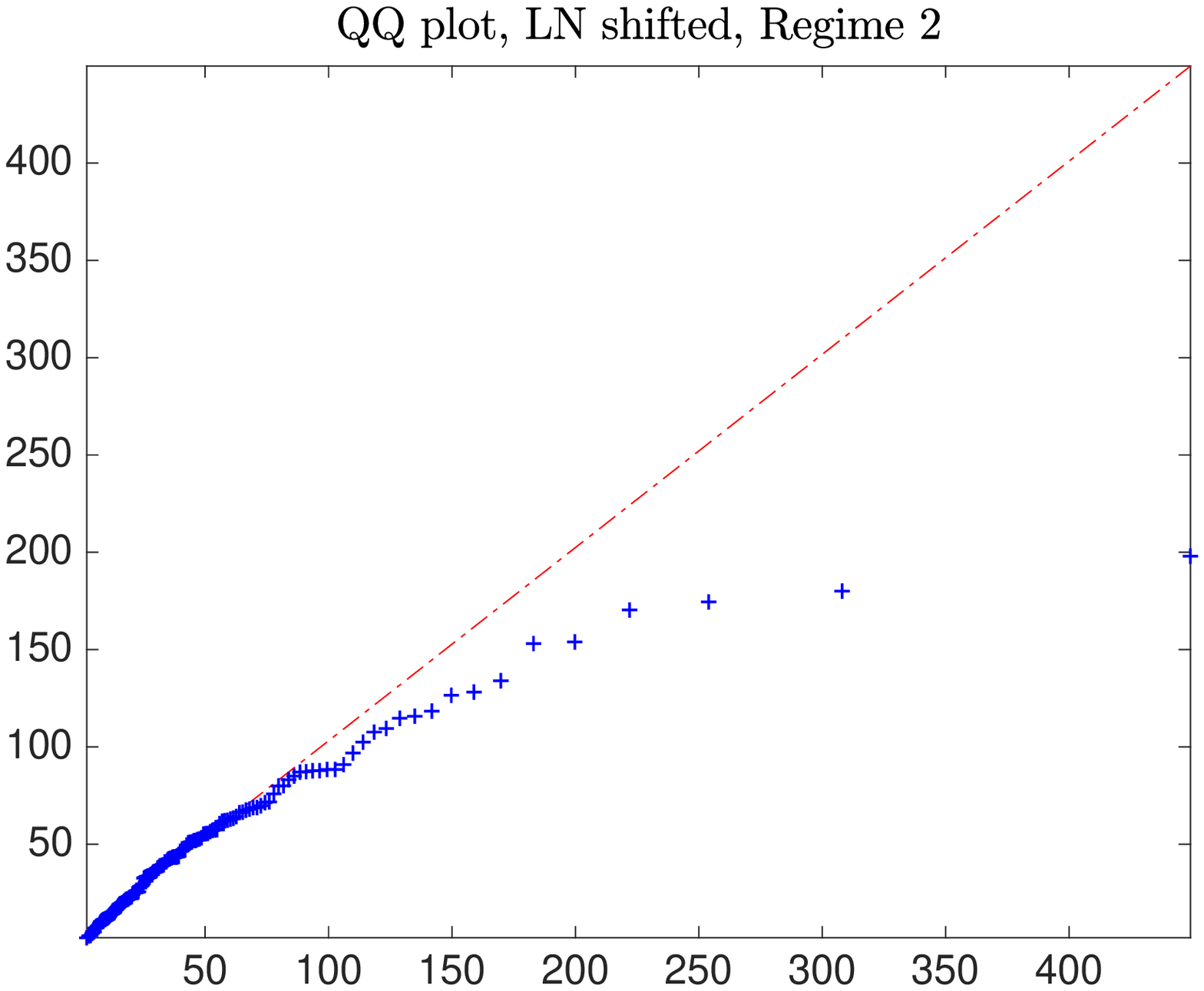}\vspace{4mm}
	\includegraphics[width = 0.5\textwidth, trim = {0 0 0 22}, clip]{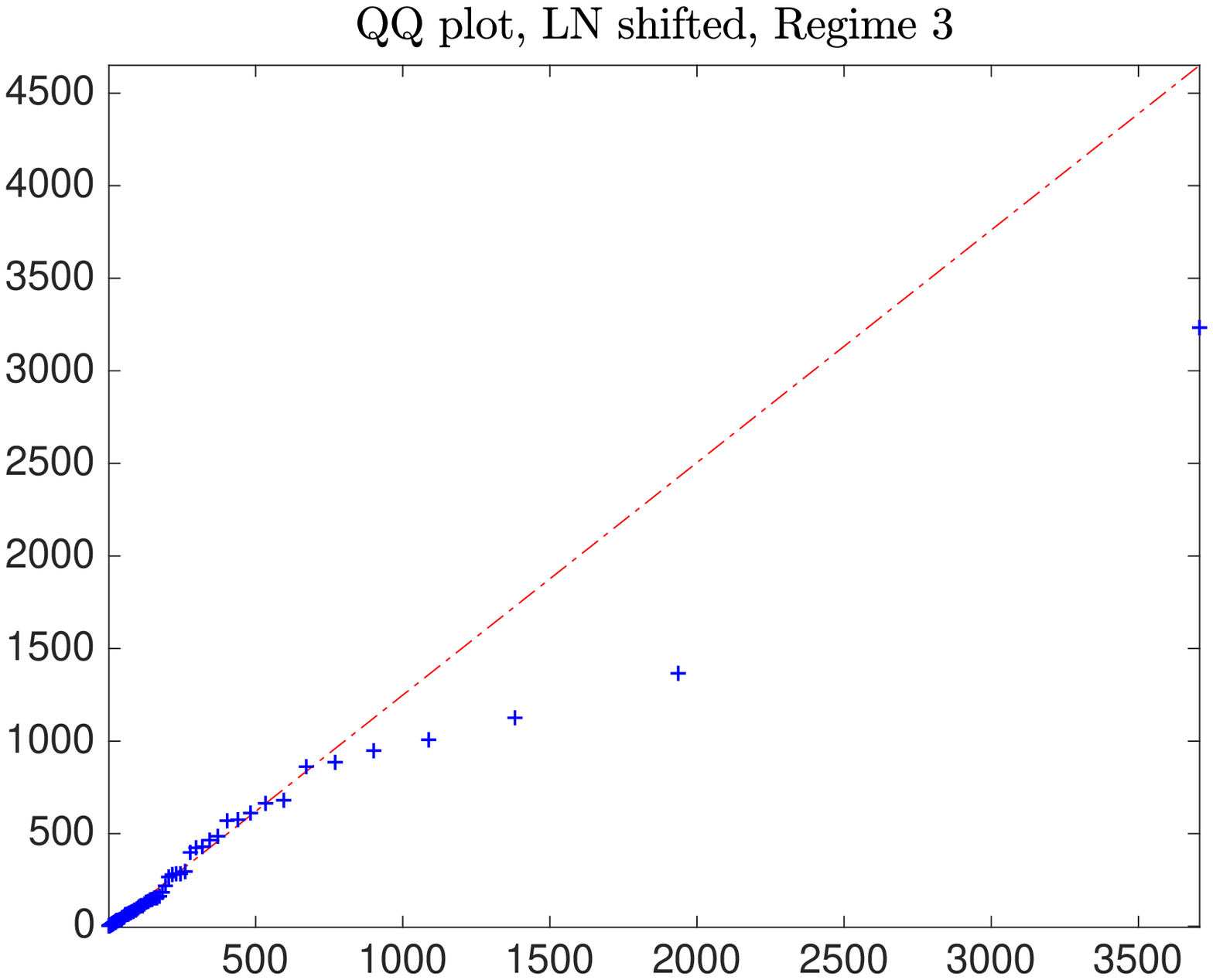}
	\caption{Representative posterior predictive check QQ plots of residuals from the posterior for \ref{eqnM2}. (Left) QQ plot of standardised residuals for the base regime, Regime~1, calculated using Equation (\ref{residenq}). (Right) QQ plot of residuals for the shifted log-normal spikes, Regime~3. (Bottom) QQ plot of the residuals of the shifted log-normal spikes, Regime~4, for extreme spikes.}
		\label{3irQQbss}
\end{figure*}

\begin{figure}
\begin{center}
\includegraphics[width = 0.5\textwidth, trim = {0 0 0 22}, clip]{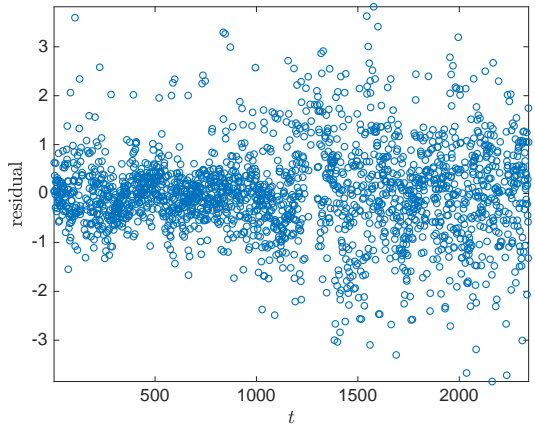}
\caption{A representative posterior predictive check plot of the residuals, calculated using Equation (\ref{residenq}), of the base Regime 1 in \ref{eqnM2}, versus time. There is a distinct jump in the variance of residuals near \(t=1200\).  }
\label{fig: residvstimeM2}
\end{center}
\end{figure}

In Figure~\ref{3irQQbss} %
representative QQ plots from our PPCs for \ref{eqnM2} are shown, which are used to assess the distributional assumptions for each regime. Figure~\ref{3irQQbss} shows that the log-normal distribution for both of the spike regimes is not reasonable -- the log-normal is too heavy-tailed. Figure~\ref{3irQQbss} also indicates that the standardised residuals of the base regime are likely not normally distributed -- the QQ plot suggests that the errors have excess kurtosis. Furthermore, as Figure \ref{fig: residvstimeM2} shows, the variance of the residuals of Regime 1 increases over time for \ref{eqnM2}. In fact there appears to be a distinct jump in the variance of the residuals around \(t=1200\) which corresponds to April 2016. One notable event that occurred around this time was the closure of South Australia's only coal generation facility \cite{abc1}. It could be that this event disrupted the market causing the jump in volatility, but more research would be needed to determine the cause of the increased volatility.%

\textbf{Model 3.} To investigate the findings from the interrogation of \ref{eqnM2}, \ref{eqnM3} is proposed, with two base regimes, to address heteroskedasticity, and two shifted log-normal spike regimes:
\begin{equation*}\label{eqnM3}
X_t = \begin{cases} 
B_t^{(1)} & \text{if } R_t = 1, \\
B_t^{(2)}& \text{if } R_t = 2,\\
Y_t^{\left(3\right)} & \text{if } R_t = 3, \\
Y_t^{\left(4\right)} & \text{if } R_t = 4. \tag{Model 3}
\end{cases}
\end{equation*}
Recall that we assume \(\sigma_1<\sigma_2\), so that the variance of base Regime~1 is less than the variance of base Regime~2. 

\begin{figure*}
	\centering	
	\includegraphics[width = 0.475\textwidth, trim = {0 0 0 22}, clip]{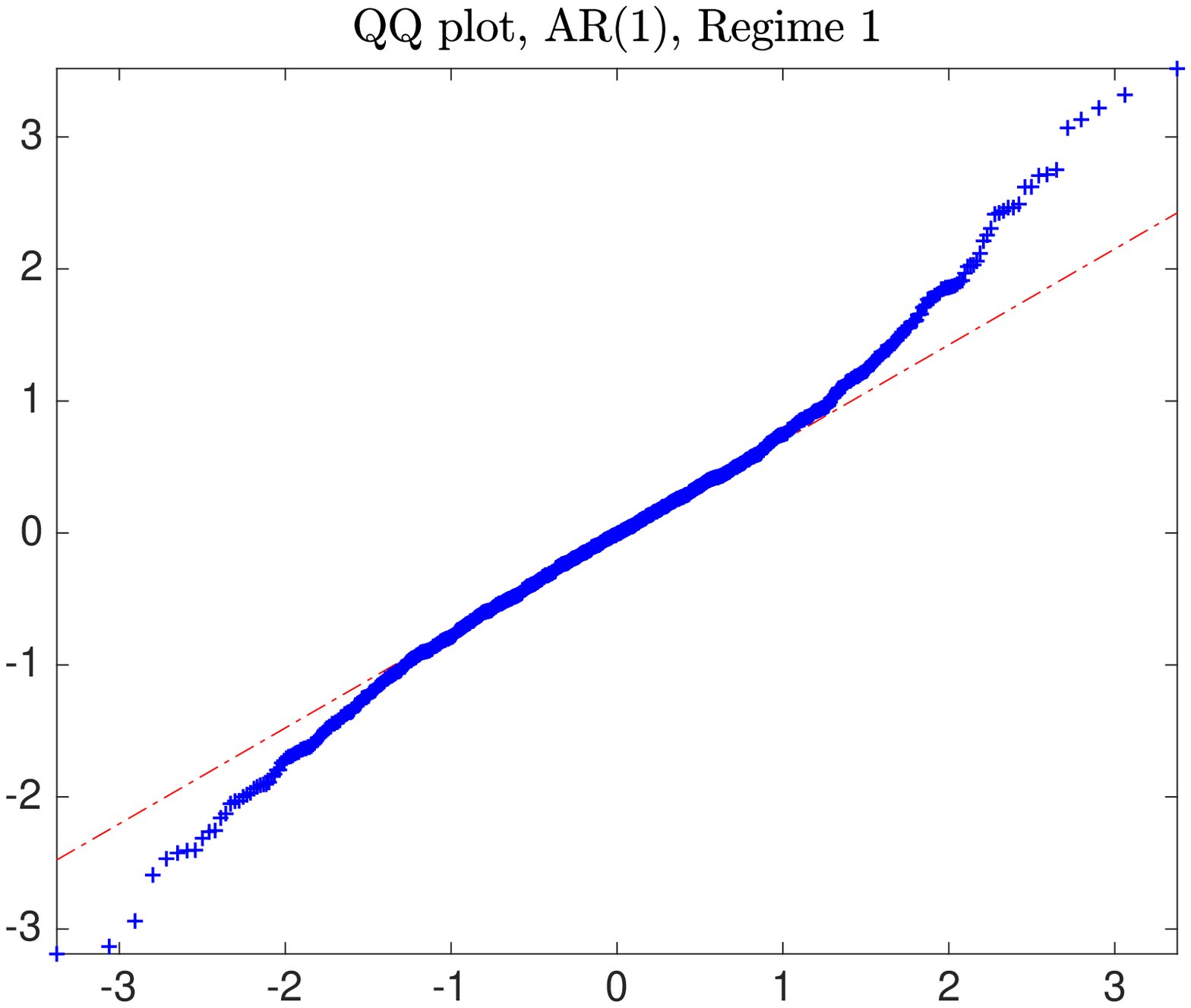}\hspace{4mm}\includegraphics[width = 0.475\textwidth, trim = {0 0 0 22}, clip]{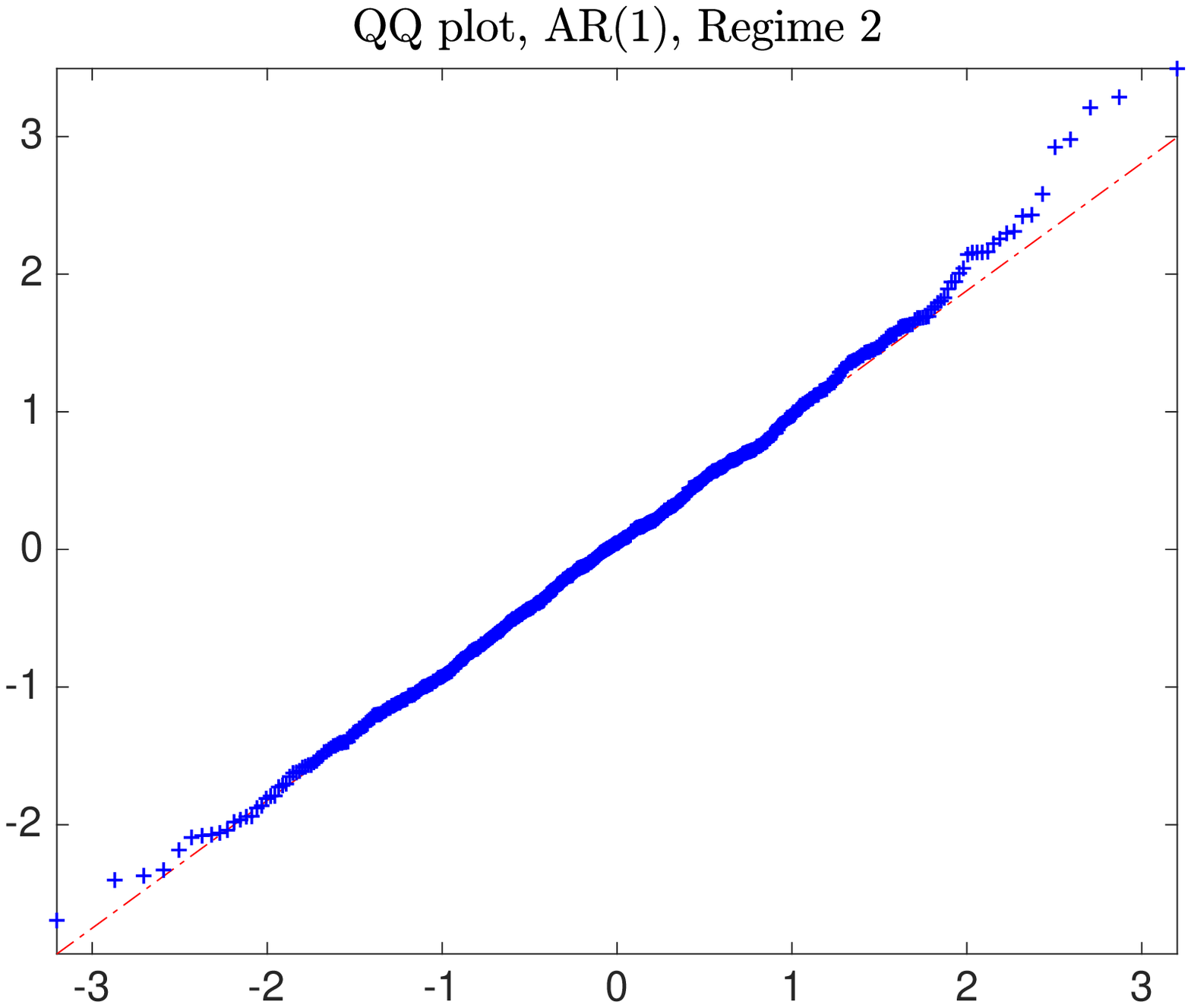}\vspace{4mm}
	\includegraphics[width = 0.475\textwidth, trim = {0 0 0 22}, clip]{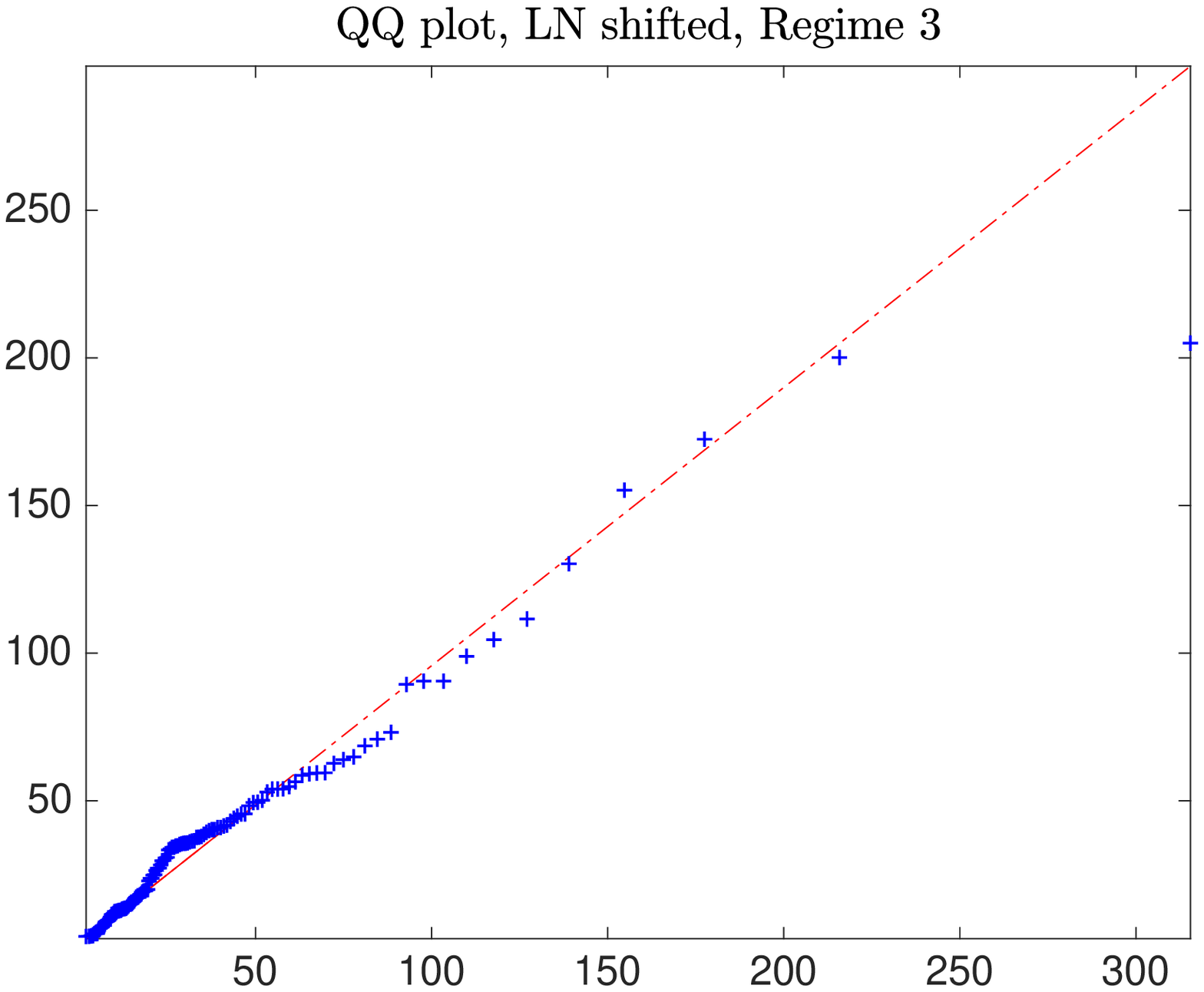}\includegraphics[width = 0.485\textwidth, trim = {0 0 0 22}, clip]{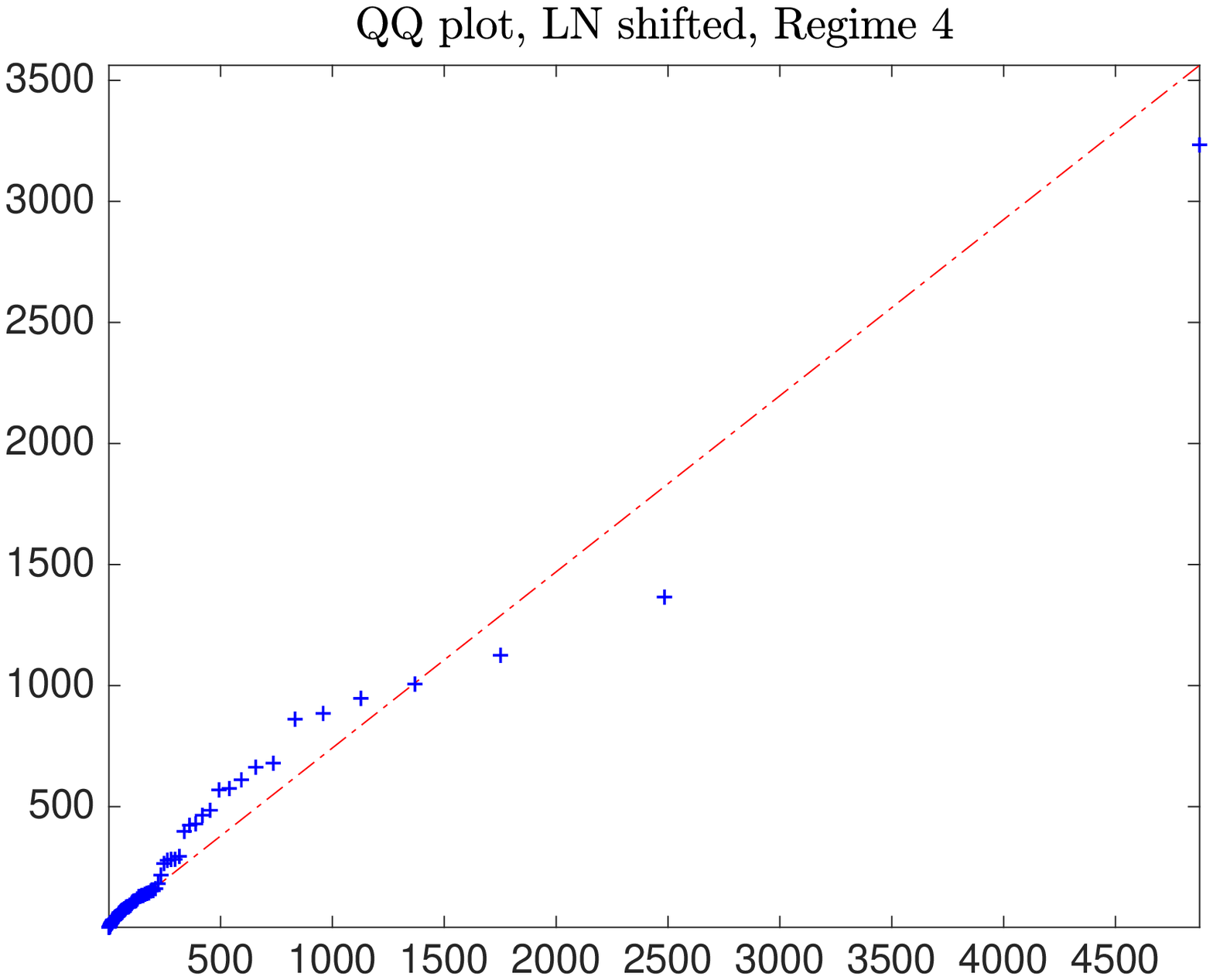}
	\caption{Representative posterior check QQ plots of residuals for a single sample of parameters and hidden regime sequence from the posterior for \ref{eqnM3}. (Top-left) QQ plot of standardised residuals of the AR(1) base Regime 1, calculated using Equation (\ref{residenq}). (Top-right) QQ plot of the standardised residuals of the AR(1) base Regime 2, calculated using Equation (\ref{residenq}). (Bottom-left) QQ plot of residuals for the shifted log-normal spike Regime~3, for typical spikes. (Bottom-right) QQ plot of the residuals of the shifted log-normal spike Regime~4, for extreme spikes. The QQ plots for Regimes 2, 3 and 4 show little curvature in the points, suggesting the distributional assumptions of these regimes are reasonable. The QQ plot for Regime 1 shows some curvature, suggesting the distributional assumptions for this regime are may still be incorrect.}
	\label{6}
\end{figure*}

Observing the QQ plots in Figure~\ref{6} that were produced as part of our PPCs, the distributional assumptions of Regimes 2, 3 and 4 are all reasonable. However the distributional assumptions of Regime 1 are questionable. There appears to be too much mass in the tails of the data compared to a normal distribution. %

\begin{figure*}
	\begin{minipage}{0.5\textwidth}
         \makebox[\textwidth][c]{\includegraphics[width = 1\textwidth, trim = {0 20 0 20}, clip]{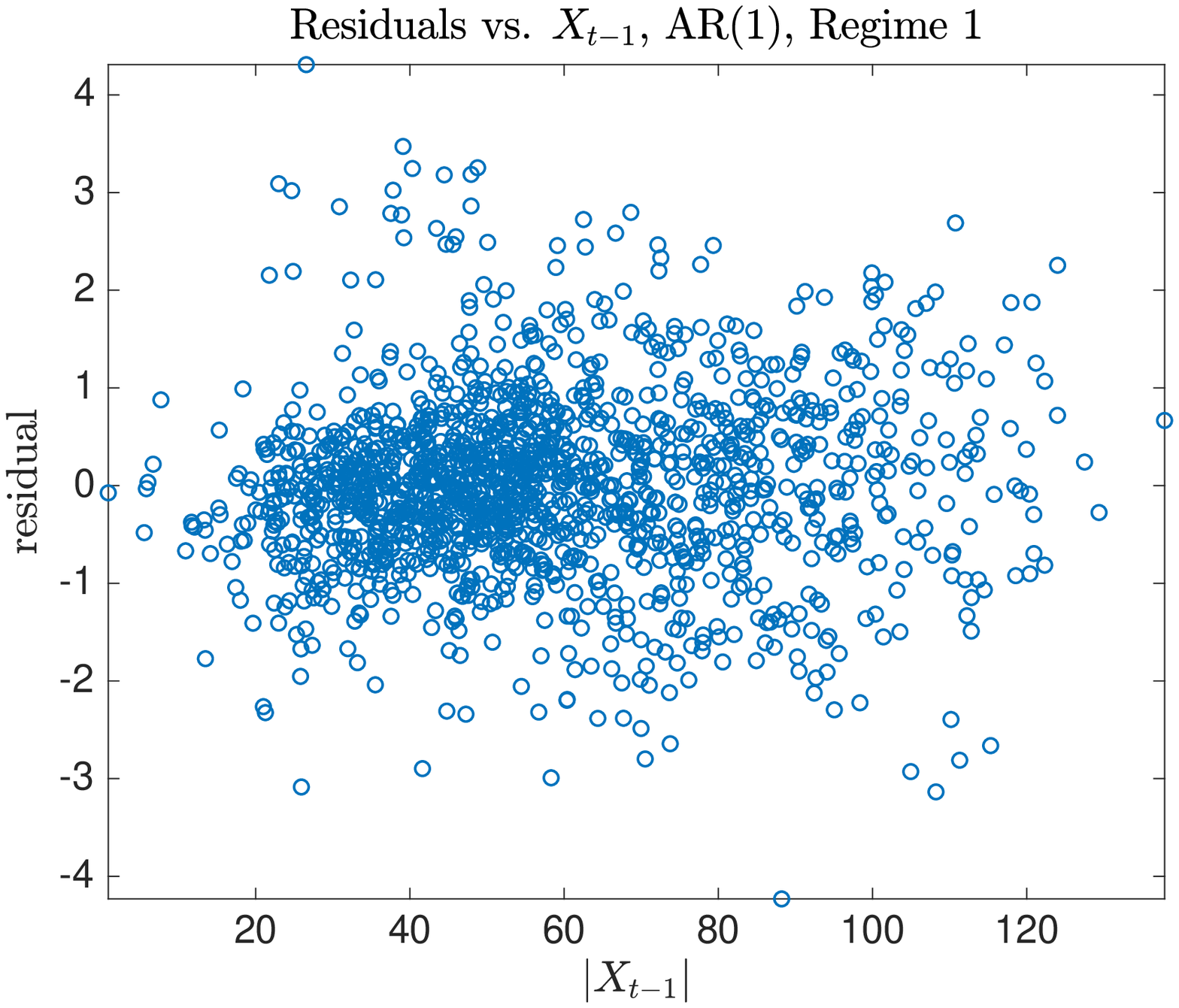}}\\
	\centering
	{\footnotesize{\(|x_{t-s_1^t}|\)}\\ \normalsize (a)}
	\end{minipage}%
	\begin{minipage}{0.5\textwidth}
         \makebox[\textwidth][c]{\includegraphics[width = 1\textwidth, trim = {0 20 0 20}, clip]{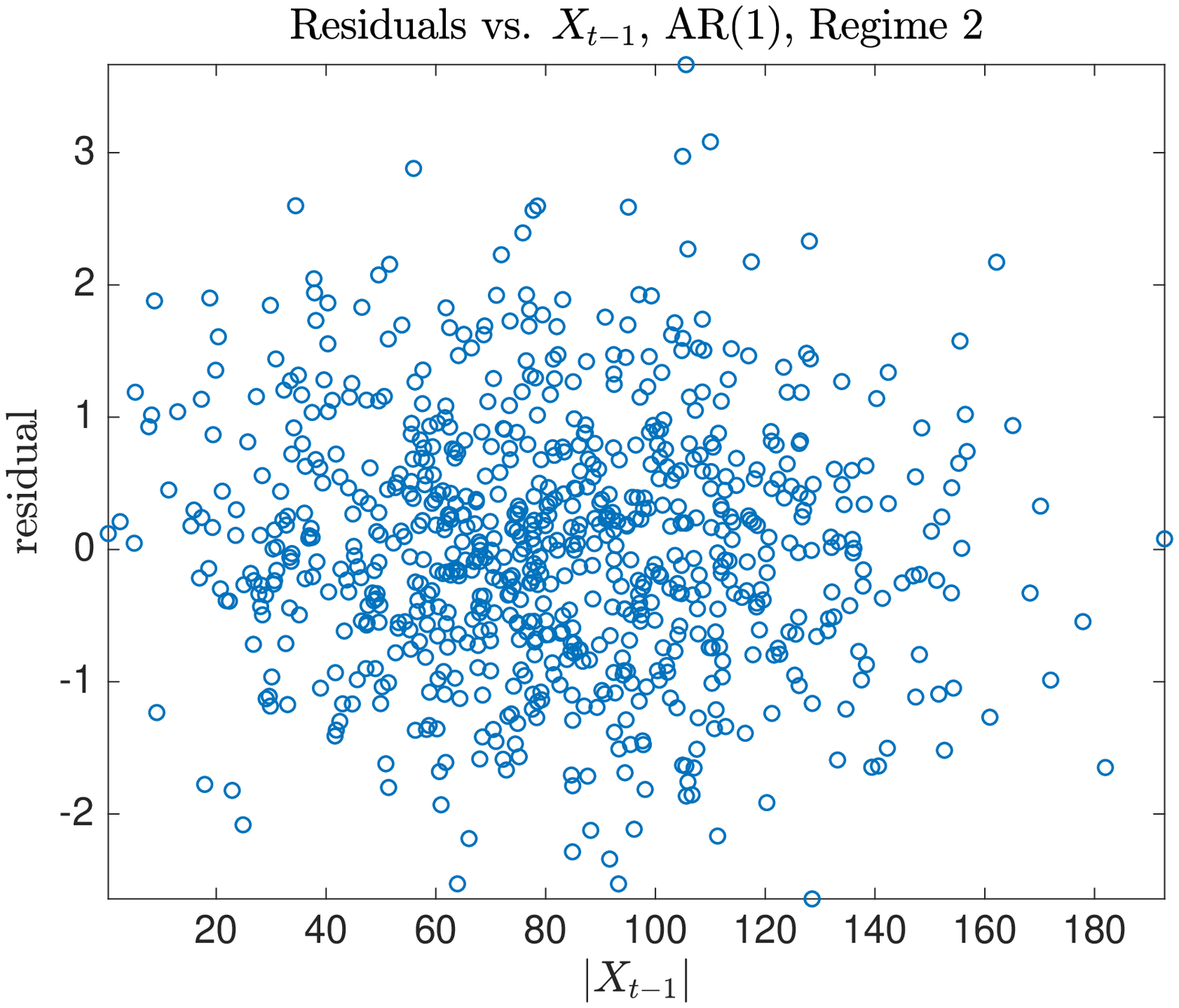}}\\
	\centering
	{\footnotesize{\(|x_{t-s_2^t}|\)}\\ \normalsize (b)}
	\end{minipage}
	\caption{Representative posterior predictive check plots of the standardised residuals of the base regimes for \ref{eqnM3}, calculated using Equation (\ref{residenq}), against the absolute value of the most recently observed lagged values, \(x_{t-s_i^t}\), of the same base regime. The lagged values \(x_{t-s_i^t}\) are determined using the sample \(\boldsymbol R^*\) from the posterior and are the last observed value before time \(t\) from (a) base Regime~1, (b) base Regime~2. In Figure (a) there is some evidence that the variance of the residuals depend on the magnitude of lagged values since the spread of the points increases with \(|x_{t-s_1^t}|\). Note that in Figure (a) there are more observed values between 20 and 70 than there are above 70 (approximately 70\% of the points lie between 20 and 70) which makes this increase in spread with \(|x_{t-s_2^t}|\) appear less pronounced than it actually is.}
	\label{residsvsxt}
\end{figure*}

To check for constant variance of residuals as a function of the most recently observed lagged value of the AR processes, PPCs are used and the standardised residuals from the base regimes are plotted against the most recently observed lagged values -- a representative plot is provided in Figure~\ref{residsvsxt} %
\begin{figure*}
    \centering
        \begin{minipage}{0.5\textwidth}
            \makebox[\textwidth][c]{\includegraphics[width = \textwidth, trim = {0 0 0 0 }, clip]{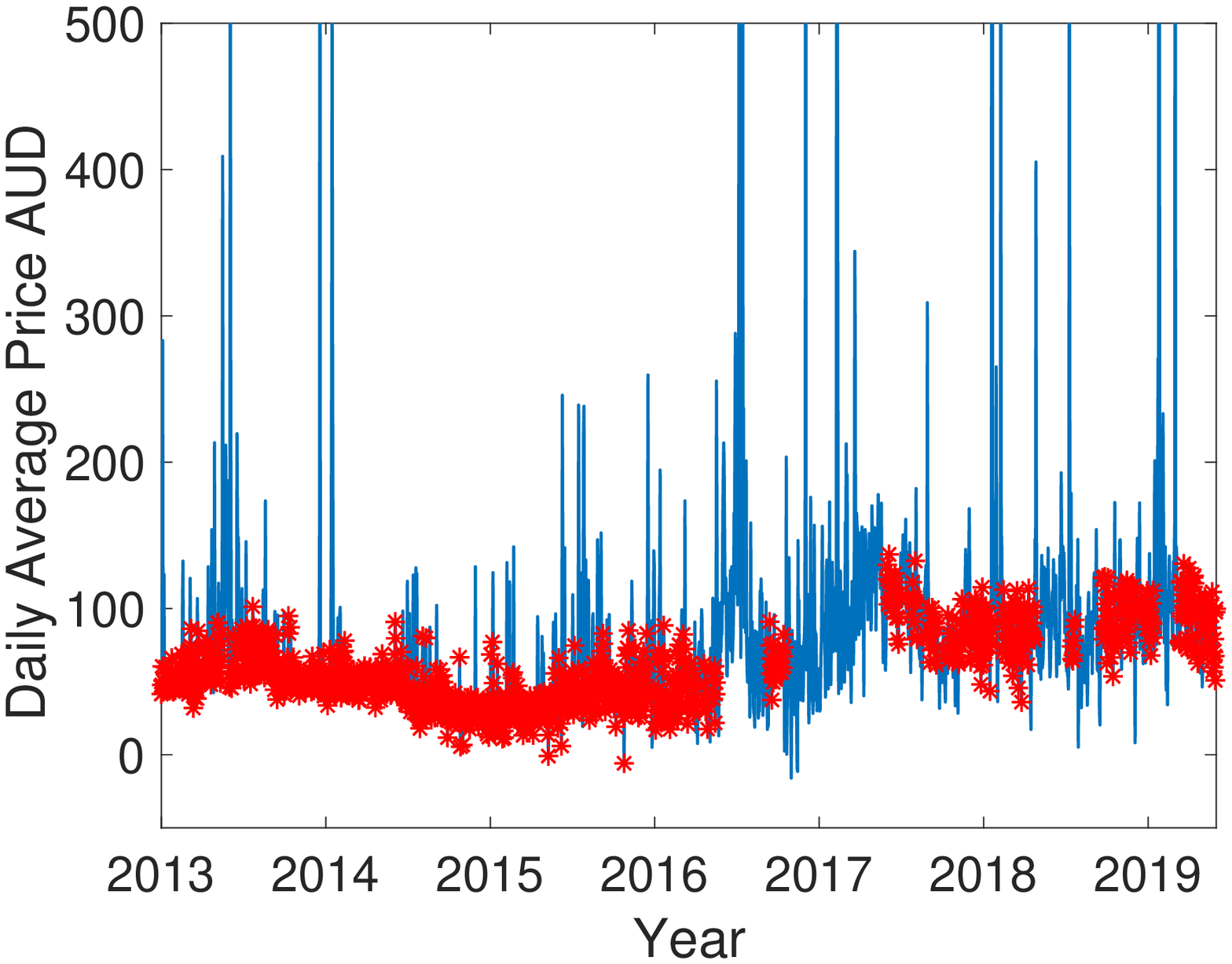}}\\
            \centering
            {(a)}
        \end{minipage}%
        \begin{minipage}{0.5\textwidth}
            \makebox[\textwidth][c]{\includegraphics[width = \textwidth, trim = {0 0 0 0}, clip]{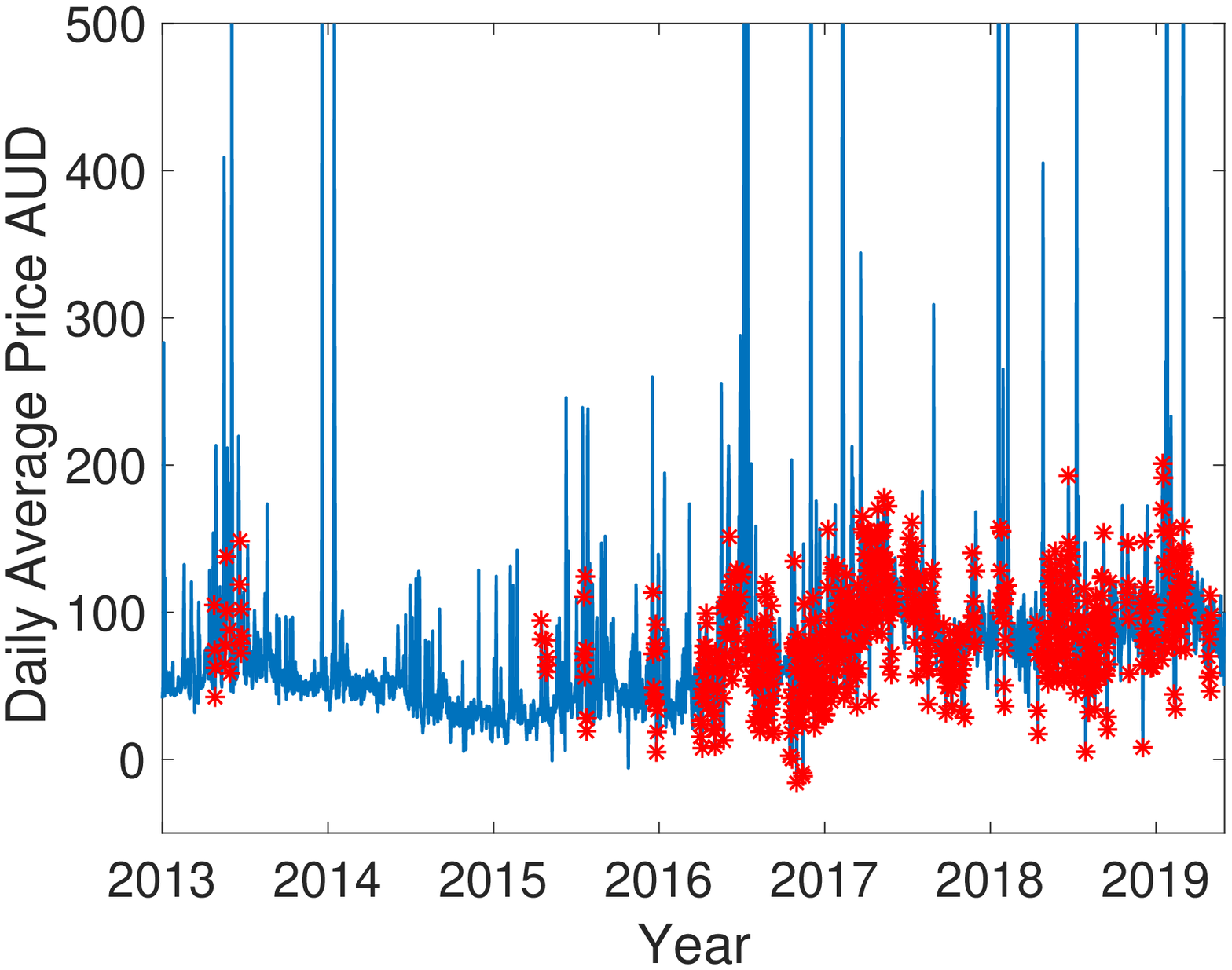}}
            \centering
            {(b)}
        \end{minipage} 
        
\vspace{1mm} 

        \begin{minipage}{0.5\textwidth}
            \makebox[\textwidth][c]{\includegraphics[width = \textwidth, trim = {0 0 0 0}, clip]{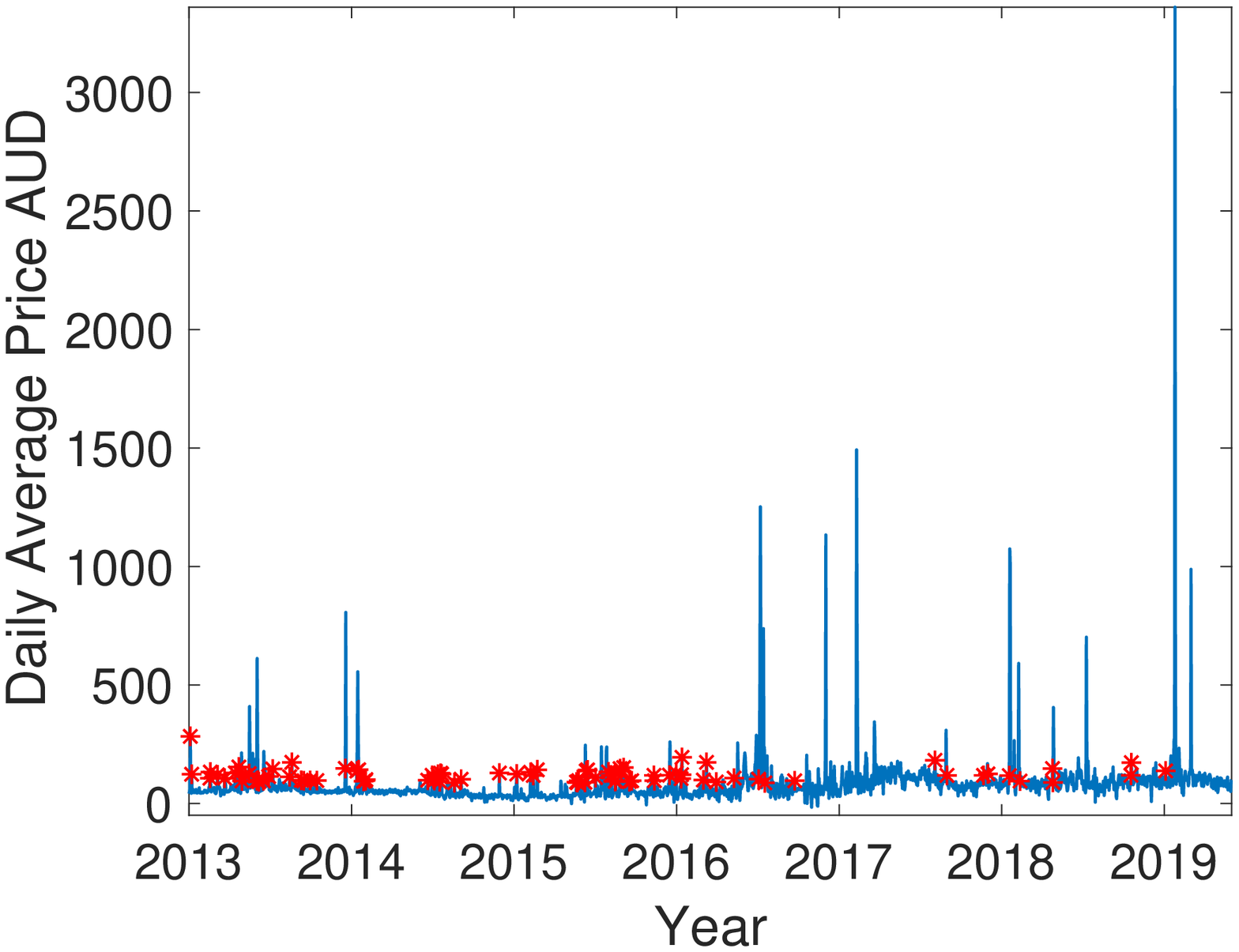}}
            \centering
            {(c)}
        \end{minipage}%
        \begin{minipage}{0.5\textwidth}
            \makebox[\textwidth][c]{\includegraphics[width = \textwidth, trim = {0 0 0 0}, clip]{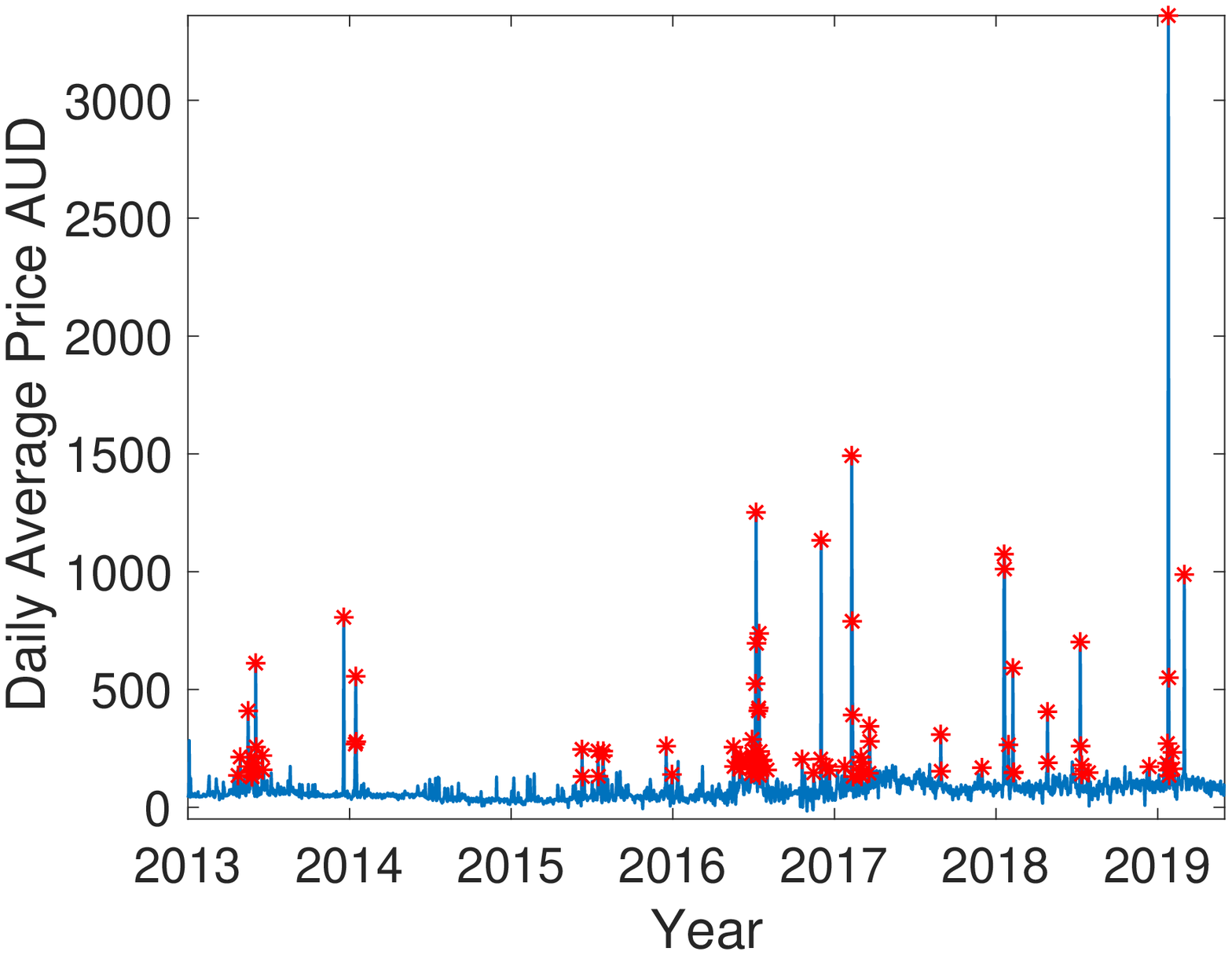}}
            \centering
            {(d)}
        \end{minipage}
    \caption{These plots display the price process, with points highlighted in red if they have the greatest (marginal) posterior probability of belonging to that regime; (a) Regime~1 -- base regime~1, (b) Regime~2 -- base regime~2, (c) Regime~3 -- spikes, (d) Regime~4 -- extreme spikes.}
    \label{4irclassification}
\end{figure*}%
for \ref{eqnM3}. In Figure \ref{residsvsxt} (a) there is some evidence that the variance depends on the magnitude of lagged values since the spread of the points increases with \(|x_{t-s_1^t}|\). In Figure \ref{residsvsxt} (b) there are no obvious signs that the variance in Regime 2 depends on the magnitude of lagged values.  

In Figure~\ref{4irclassification} we plot the price series and a classification of prices into regimes according to \ref{eqnM3}. We classify prices into Regime \(i\) if the posterior probability of the price being generated by Regime \(i\) in \ref{eqnM3} is greatest of all the regimes. That is, we classify in to Regime \(i\) if \(i = \mbox{arg}\max_{j}P(R_t=j|\boldsymbol x)\). In Figure \ref{4irclassification} we see that the low-volatility base regime, Regime~1, mostly captures base prices from January 2013 to April 2016, and the high-volatility base regime, Regime~2, mostly captures base prices from April 2016 onwards. Thus, \ref{eqnM3} captures the jump in market volatility in 2016 with Regime 2. Previously non-constant variance of the base regime has been modelled by a \emph{constant elasticity of volatility} (CEV) process, \(B_t=\phi B_{t-1} + \sigma |B_{t-1}|^\gamma\varepsilon_t\) \cite{janczura2009}. However, this type of model is not an obvious candidate to capture the jump in volatility here since the jump in volatility is more persistent than we would expect from the CEV model. Although, the CEV model could be used to rectify the excess mass in the tails of the distribution of the residuals for Regime 1 in \ref{eqnM3}. Figure \ref{4irclassification} also illustrates that the regime process, \(\{R_t\}\), may not be time-homogeneous, or even Markovian, for \ref{eqnM3}.

In Table \ref{table: parameters} we show the posterior mean estimates for the parameters of the models discussed above, as well as the natural logarithm of the Bayes factor with reference to \ref{eqnM3}. The Bayes factor provides decisive evidence that \ref{eqnM3} is favourable over the others. 

\begin{table}[h!]
\caption{Posterior mean parameter estimates. }
\centering
\begin{tabular}{ @{\hskip 0in} l @{\hskip 0in} | l | @{\hskip 0in} c @{\hskip 0in} c @{\hskip 0in} c @{\hskip 0in}} 
&&\ref{eqnM1} & \ref{eqnM2} & \ref{eqnM3}\\ \hline
\multirow{2}{*}{\begin{tabular}{@{\hskip 0in} c@{\hskip 0in} }Base \\regime 1\end{tabular} }&\(\phi_1\) & 0.369 & 0.392 & 0.280 \\ 
&\(\sigma_1^2\) & 351 & 306 & 153 \\ \hline
\multirow{2}{*}{\begin{tabular}{@{\hskip 0in} c@{\hskip 0in} }Base \\regime 2\end{tabular} }&\(\phi_2\) & -& -& 0.585 \\ 
&\(\sigma_2^2\) & -& -& 756 \\ \hline
\multirow{3}{*}{\begin{tabular}{@{\hskip 0in} c@{\hskip 0in} }Spike \\regime 1\end{tabular} }&\(q_3\) & 85.1 & 83.1 & 84.0 \\ 
&\(\mu_3\) & 4.07 & 3.66& 3.22 \\ 
&\(\sigma_3^2\) & 1.35 & 0.619 & 0.873 \\ \hline
\multirow{3}{*}{\begin{tabular}{@{\hskip 0in} c@{\hskip 0in} }Spike \\regime 2\end{tabular} }&\(q_4\) & -& 135 & 126 \\ 
&\(\mu_4\) & -& 4.58 & 4.26 \\ 
&\(\sigma_4^2\) &-& 2.56 & 2.45 \\ \hline
\multirow{3}{*}{\begin{tabular}{@{\hskip 0in} c@{\hskip 0in} }Drop \\regime\end{tabular} }&\(q_5\) & -50.2 & -& -\\ 
&\(\mu_5\) & 2.97 & -& -\\ 
&\(\sigma_5^2\) & 0.120 & -& -\\\hline
\begin{tabular}{@{\hskip 0in} c@{\hskip 0in} }Regime \\ probs.\end{tabular} & \(p_{ij}\) & \(\left[\begin{array}{ccc}
    .943  &  .051  &  .006\\
    .407  &  .578  &  .015\\
    .057  &  .018  &  .925
    \end{array}\right]\) & 
    \(\left[\begin{array}{ccc}
    .942  &  .053  &  .005\\
    .527  &  .335  &  .137\\
    .167  &  .344  &  .489
    \end{array}\right]\)
    & \(\left[\begin{array}{cccc}
    .934&    .014&    .050&    .002\\
    .028&    .916&    .010&    .046\\
    .664&    .042&    .154&    .140\\
    .053&    .350&    .100&    .497
    \end{array}\right]\) \\ \hline
    \begin{tabular}{@{\hskip 0in} c @{\hskip 0.01in} }log-Bayes \\factor\end{tabular} & & 149 & 159 & 0
 \end{tabular}\label{table: parameters}
\end{table}

For completeness, we also fitted a model with all 5 regimes but found that the drop regime was completely unnecessary since no points were classified as drops. Moreover, the Bayes factor for the 5 regime model also provided strong evidence that the drop regime was unnecessary. The natural log of the Bayes factor between \ref{eqnM3} and the same model with an added drop regime was 13.07. 

We also fit \ref{eqnM3} with shifted gamma spike regimes instead of log-normal. For this model the PPCs (not shown) showed little evidence of an ill-fitting model; they were comparable to the PPCs of \ref{eqnM3}. However, the Bayes factor suggested that log-normally distributed spikes were better. The natural log of the Bayes factor of the model with gamma spikes compared to \ref{eqnM3} was 8.06.

\paragraph{Discussion: All models are wrong, some are useful -- George Box}
Here we have used PPCs as well as a classification of prices into regimes from models to interrogate models and assess in which ways they fail to fit the data. Using these methods we are able to clearly see how models may misrepresent the data. For example, we are able to recognise that the models used in this paper are unable to completely capture distributional properties of base prices. Furthermore, our PPCs eluded to the fact that there is a distinct change-point in volatility in the data, which led us to include a regime to capture this feature. Thus PPCs are an extremely valuable tool to inform us about which aspects of the models are wrong and in which ways. Hence we may make more informed decisions about the conclusions we draw from our modelling, or how we use the models. We have also used Bayes factors to compare models. Statistical summaries of model such as Bayes factors are useful metrics for model comparison, but unlike PPCs, give no indication if models accurately capture the features of the data. 

This modelling highlights another issue, which is that electricity markets are dynamic; they are constantly changing. For example there is an increasing push for renewable resources, and less reliance on traditional fossil fuels; there are regular policy and regulatory changes; and there are systematic influences from other commodity markets, such as gas markets. As a result, it is hard to model these markets with models which cannot account for these changes. There has been some work to reconcile this, see \cite{becker2007,noren2013} for example, where exogenous factors are included in MRS models, to capture changing market conditions. The novel methods presented in this paper -- the PPCs, the Bayesian approach including the use of MCMC, and the integrated trend modelling process -- can be used to build models with exogenous factors and infer their parameters. We see this as an interesting direction for future research.

We should also look to challenge other modelling assumptions, such as the time-homogeneous Markovian nature of the occurrence of price spikes. The occurrence of price spikes as a point process has been studied, see \cite{becker2013,christensen2012,christensen2009,handika2014,huisman2008}, for example, but these results have not been used to inform MRS models for electricity prices. We believe that the methodologies presented in this paper can be used to develop non-time-homogenous MRS models and to include exogenous factors in the switching parameters. We suggest this as an area for further research for the electricity price modelling community.

\section{Conclusions}\label{conclusion}
In this paper we have discussed and extended the methodologies for stochastic MRS and trend models for wholesale electricity prices. Trend modelling for electricity prices is not straightforward since large price spikes can skew trend estimates. We suggest that the trend model is integrated with the stochastic model and all components estimated jointly. By modelling the trend in this way we are able to stop large price spikes affecting trend estimates since the MRS model is able to identify spikes in a very natural way as part of the estimation procedure. Furthermore, combining the trend and stochastic models permits the use of statistical model selection criteria, such as the BIC and Bayes factors, for both the trend and stochastic components. We have also shown how Bayes factors and posterior predictive checks can be used to inform modelling choices and for model comparisons. We have shown how to include two different types of trend models within the stochastic component; a wavelet-based model and a cubic splines-based model. Fitting the models to the South Australian data we found that the choice of trend model did not greatly affect the ultimate conclusions. Therefore, due to the relative simplicity we recommend the cubic spline model over the wavelet-based model. 

We have also shown how to estimate the model in a Bayesian setting. For electricity price modelling, model estimation is typically done via an approximation to the EM algorithm. By using a Bayesian approach we are able to estimate model parameters exactly. Furthermore, a Bayesian approach to estimation of shifting parameters of spike distributions can alleviate some issues regarding identifiability of the model in the maximum-likelihood context. This approach also allows us to easily implement posterior predictive checks so assess model fit. 

The methods presented in this paper were demonstrated by an application to the South Australian wholesale electricity market. We found that the cubic-spline based trend model was favourable over the wavelet-based model. We also found that no drop regime was necessary; that two spike regimes may be required, one to capture a typical spike, and one to capture the very extreme observations; and that there was a significant jump in volatility in prices in 2016, which could not be explained by the stationary base regimes. More work is required to be able to adequately capture this feature, and we suggest including exogenous factors in the model as a way to do this. The methodologies developed in this paper can be extended to include exogenous factor in the model.

\section*{Acknowledgements}
All authors would like to acknowledge the financial support of the Australian Research Council Centre of Excellence in Mathematical and Statistical Frontiers (ACEMS), and the first author would also like to acknowledge the support of the Australian government Research Training Program (RTP) and The University of Adelaide Master of Philosophy (No Honours) scholarship.

\FloatBarrier

\bibliography{mybib}

\clearpage
\section{Supplementary material}
\subsection{Wavelet-based trend model}

\paragraph{Preliminaries}
Wavelet filtering is based on \emph{multiresolution analysis}, where a function (or signal, or time series) is decomposed and represented as a sum of functions at various levels of resolution. That is, we have a function \(f(x)\) that we can represent as 
\[f(x) = \sum_{j=-\infty}^\infty \sum_{k=-\infty}^\infty b_{j,k}\psi_{j,k}(x),\]
where \(b_{j,k}\), \(j,k\in \mathbb Z\), are coefficients, and \(\psi_{j,k}\), \(j,k\in \mathbb Z\), are \emph{translated} and \emph{dilated} \emph{wavelet} functions. For fixed \(j\), we may think of the terms involving the functions \(\psi_{j,k}\), \(k\in \mathbb Z\), as representing the features of the function, \(f\), at resolution \(j\). We define the wavelet functions \(\psi_{j,k}\), \(j,k\in \mathbb Z\), from a single \emph{mother wavelet}, \(\psi(x)\) by translation, shifting the function by integers \(k\in\mathbb Z\), and dilation, scaling the function by \(2^j\),
\[\psi_{j,k}(x) = 2^{j/2}\psi(2^jx-k).\]

To form a multiresolution analysis, we require a sequence of subsets \(\{V_j\}\) such that 
\[\{0\} ... \subset V_{-1}\subset V_0\subset V_1 \subset ... \subset L^2,\]
where \(L^2\) is the set of \emph{square integrable} functions (the set of functions, \(g\), such that \(\int_{\mathbb R} |g(x)|^2 dx<\infty\)). Moreover, we also require \(\bigcup\limits_{j\in\mathbb Z} V_j\) to be dense in \(L^2\), \(\bigcap\limits_{j\in\mathbb Z}V_j = \{0\}\), the scaling property implies \(g(x) \in V_j \iff g(2x) \in V_{j+1}\), and we must also have the closure under translation property \(g(x)\in V_j \iff g(x+2^{-j}) \in V_j\). 

We also need a set of orthonormal basis functions, \(\phi(x-k)\), \(k \in \mathbb Z\), called scaling functions, which form a basis for \(V_0\), and therefore, \(\phi_{j,k}(x) = 2^{j/2}\phi(2^jx-k)\), \(k\in\mathbb Z\) is a basis for \(V_j\), for each \(j\in\mathbb Z\). Since \(V_0\subset V_1\), then 
\begin{equation}\label{eqn: h's}\phi(x) = \sum_{k=-\infty}^\infty h_k \phi(2x-k).\end{equation}
Here we assume there are only a finite number, \(N\), of non-zero terms in \(\{h_k,k\in\mathbb Z\}\), namely \(h_0>0,...,h_{N-1}>0\), which is the case for the scaling function related to the Daubechies wavelets. 

The Daubechies wavelet family is a family of wavelet functions indexed by the number of \emph{vanishing moments} which they possesses. A wavelet has \(p\) vanishing moments if \(\int_{\mathbb R} \psi(x)x^\ell = 0\) for \(\ell = 0,1,...,p-1\). This property means that any function of the form \(f(x)=\alpha_0+\alpha_1x+\dots+\alpha_{p-1}x^{p-1}\) can be exactly represented as \(f(x) = \sum\limits_{k=-\infty}^\infty a_{0,k}\phi(x)\). 

We can also characterise the difference between the subspaces \(V_j\) and \(V_{j+1}\). Let \(W_j\) be the \emph{orthogonal complement} of \(V_j\) in \(V_{j+1}\). Then \(W_j\), \(j\in\mathbb Z\), are orthogonal and the \emph{direct sum}, \(\bigoplus\limits_{j\in\mathbb Z} W_j = L^2\). Now let \(\psi(x-k)\), \(k\in\mathbb Z\), be an orthonormal set of basis functions for \(W_0\). Therefore \(\psi_{j,k}=2^{j/2}\psi(2^jx-k)\), \(k\in\mathbb Z\), is an orthonormal basis for \(W_j\). Since \(W_0 \in V_1\), then \(\psi\) can be represented as 
\begin{align}\label{eqn: high pass}\psi(x) = \sum_{k=-\infty}^\infty g_k\phi(2x-k).\end{align}
The function \(\psi\) is known as the mother wavelet, and \(\psi_{j,k}\) are wavelet functions. 

Moreover, \(\psi_{j,k}(x)\), \(j,k\in\mathbb Z\) is an orthonormal basis for \(L^2\). That is, any function \(f\) in \(L^2\) can be represented as 
\[f(x) = \sum_{j=-\infty}^\infty \sum_{k=-\infty}^\infty b_{j,k}\psi_{j,k}(x),\]
where the coefficients \(b_{j,k} = \int_{\mathbb R} f(x)\psi_{j,k}(x)dx\), are the inner product of \(f\) and \(\psi_{j,k}\). 

\paragraph{Approximation} A smoothed approximation to \(f\) may be obtained by removing all terms above resolution \(M-1\). That is, a smoothed approximation at resolution \(M\), \(\widehat f_M\), to \(f\), is 
\begin{equation}\label{eqn: wavelet approx}\widehat{f}_M(x) = \sum_{j=-\infty}^{M-1} \sum_{k=-\infty}^\infty b_{j,k}\psi_{j,k}(x).\end{equation}
Since \(\bigoplus\limits_{\{j\mid j\in Z, j < M\}}W_j \subset V_{M}\), then Equation (\ref{eqn: wavelet approx}) may also be written in terms of scaling functions;  
\begin{equation}\label{eqn: scaling approx}\widehat{f}_M(x) = \sum_{k=-\infty}^\infty a_{M,k}\phi_{M,k}(x),\end{equation}
where \(a_{M,k}\) is the inner product of \(f\) and \(\phi_{M,k}\), \(a_{M,k} = \int_{\mathbb R} f(x)\phi_{M,k}(x)dx\). These are known as the \emph{approximation} coefficients. 

\paragraph{Decomposition and reconstruction} Due to the fact that \(W_{M-1}\) and \(V_{M-1}\) are orthogonal and \(W_{M-1} \bigoplus V_{M-1} = V_{M}\), then we may decompose \(\widehat f_M\) in terms of the bases for \(W_{M-1}\) and \(V_{M-1}\); 
\begin{align}\widehat{f}_M(x) &= \sum_{k=-\infty}^\infty a_{M,k}\phi_{M,k}(x) \nonumber\\&= \sum_{k=-\infty}^\infty a_{M-1,k}\phi_{M-1,k}(x) + \sum_{k=-\infty}^\infty b_{M-1,k}\psi_{M-1,k}(x) \nonumber
\\& =\widehat f_{M-1}(x) + \sum_{k=-\infty}^\infty b_{M-1,k}\psi_{M-1,k}(x), \label{eqn: decomp intuit}
\end{align}
where \(a_{M-1,k} = \int_{\mathbb R} f(x)\phi_{M-1,k}(x)dx\) and \(b_{M-1,k} = \int_{\mathbb R} f(x)\psi_{M-1,k}(x)dx\). Rearranging gives 
\begin{align}\label{eqn: decomp approx} \widehat f_{M-1}(x) = \widehat f_M(x) - \sum_{k=-\infty}^\infty b_{M-1,k}\psi_{M-1,k}(x)\end{align}
Due to orthogonality properties we may write \(a_{M-1,k} = \int_{\mathbb R} \widehat f_{M-1}(x)\phi_{M-1,k}(x)dx\), and upon substituting Equation (\ref{eqn: decomp approx}) we find that 
\begin{equation}\label{eqn: a decomp}a_{M-1,k} = 2^{-1/2} \sum_{j=-\infty}^\infty a_{M,j}h_{j-2k},\end{equation}
where \(\{h_k\}\) are the same as those in Equation (\ref{eqn: h's}). Similarly, we find a relation for \(b_{M-1,k}\) in terms of \(a_{M,k}\); 
\begin{equation}\label{eqn: b decomp}b_{M-1,k} = 2^{-1/2} \sum_{j=-\infty}^\infty a_{M,j}(-1)^jh_{N-1-j+ 2k}.\end{equation}
It can be shown that the filter coefficients which appear in Equation (\ref{eqn: high pass}), are related by \((-1)^jh_{N-1-j+ 2k}=g_k\).

Similar formulas for reconstructing coefficients \(\{a_{M,k},k\in\mathbb Z\}\) from \(\{a_{M-1},k\in\mathbb Z\}\) and \(\{b_{M-1,k}\in\mathbb Z\}\) can be derived. Using Equation (\ref{eqn: decomp intuit}) in \(a_{M,k}=\int_{\mathbb R} \widehat f_M(x)\psi_{M,k}(x)dx\), we find 
\begin{align}\label{eqn: reconstruction} a_{M,k}=2^{-1/2}\sum_{\ell=-\infty}^\infty a_{M-1,\ell}h_{k-2\ell}+2^{-1/2}\sum_{\ell=-\infty}^\infty b_{M-1,\ell}(-1)^kh_{N-1-k+2\ell}.\end{align}

\paragraph{Discrete data and initialisation}
Let \(\boldsymbol x = \{x_k\}_{k=-\infty}^\infty\) be a discretely observed process that we want to approximate using wavelet filtering. We can consider the observations \(x_n\) as averages of a function \(f\) over the intervals \((n-\frac{1}{2},n+\frac{1}{2})\). To proceed, we need to find a suitable function \(f\) with the property 
\[x_n=\int_{n-\frac{1}{2}}^{n+\frac{1}{2}}f(x)dx,\]
and then compute \(a_{N,k} = \int_{\mathbb R} f(x)\phi_{N,k}(x)dx\), then we may proceed with the decomposition algorithm. Theoretically, (under certain conditions), the correct \(a_{N,k}\) can be found, however, here we use a simpler and faster approximation method \cite{stephane2009}. Consider 
\[f(x) = \sum_{k=-\infty}^\infty x_k \phi_{N,k}(x).\]
Therefore 
\begin{align}
a_{N,m}&=\int_{\mathbb R} f(x)\phi_{N,m}(x)dx \nonumber
\\&=\int_{\mathbb R} \sum_{k=-\infty}^\infty x_k \phi_{N,k}(x)\phi_{N,m}(x)dx \nonumber
\\&= x_m.\label{eqn: init}
\end{align}
Also, since \(\int_{\mathbb R}\phi_{N,m}(x)dx = 1\),
\begin{align*}
a_{N,m}&=\int_{\mathbb R} f(x)\phi_{N,m}(x)dx
\end{align*}
can be viewed as a weighted average of \(f(x)\) over the support of \(\phi_{N,m}\). Hence, assuming \(f\) is nicely behaved we can approximate \(x_m\approx f\) on the support of \(\phi_{N,m}(x)\). 

\paragraph{Decomposition and reconstruction as matrix operations on discrete data} To approximate (smooth) a time series by removing the high resolution (frequency) components of the data, the relations in Equations (\ref{eqn: a decomp}) and (\ref{eqn: b decomp}), along with the initialisation in Equation (\ref{eqn: init}), suggest a sequential approximation procedure. We may represent the sequential procedure by matrix operations, and this will help to illuminate the equivalence between wavelet filtering and ordinary least squares regression. 

Assume a doubly-infinite signal \(\boldsymbol x=(\dots,x_{t-1},x_t,x_{t+1},\dots)\), and let \(H\) be the infinite dimensional matrix 
\begin{align*}
H=
2^{-1/2}\left[\begin{array}{ccccccccccc}
\ddots&\ddots&\ddots&&\ddots&&&&\\
0&h_0&h_1&h_2  & \hdots & h_{N-1} &0& 0&0\\
0&0&0&h_0& \hdots & h_{N-3} & h_{N-2}&h_{N-1} & 0\\
&&&&&&\ddots&\ddots&\ddots
\end{array}\right],
\end{align*}
and denote \(\boldsymbol a_j = (\hdots,a_{j,k-1},a_{j,k},a_{j,k+1},\hdots)'\). Note that \(H\) has the property \(H'H=I\), the identity. 
Then, to find the coefficients \(\boldsymbol a_{N-J}\), we initialise the algorithm with \(a_{N,k} = x_k\) and compute
\[\boldsymbol a_{N-J} = H\boldsymbol a_{N-J+1} = H^J\boldsymbol a_{N} = H^J\boldsymbol x.\] 

Defining \(G\) as 
\begin{align*}
G=
2^{-1/2}\left[\begin{array}{ccccccccccc}
\ddots&\ddots&\ddots&&\ddots&&&&\\
0&g_0&g_1&g_2  & \hdots & g_{N-1} &0& 0&0\\
0&0&0&g_0& \hdots & g_{N-3} & g_{N-2}&g_{N-1} & 0\\
&&&&\ddots&&\ddots&\ddots&\ddots
\end{array}\right],
\end{align*}
we may similarly calculate the coefficients \(\boldsymbol b_j = (\hdots,b_{j,k-1},b_{j,k},b_{j,k+1},\hdots)'\), via
\[\boldsymbol b_{N-J} = G\boldsymbol a_{N-J+1} = GH^{J-1}\boldsymbol a_{N} = GH^{J-1}\boldsymbol x.\] 

Reconstruction can also be written in terms of the operators \(G\) and \(H\);
\[\boldsymbol a_{M} = H'\boldsymbol a_{M-1} + G'\boldsymbol b_{M-1}.\]

This reconstruction formula can be used to find an estimate of the smoothed signal. Let \(\widehat f_{N-J}\) be an approximation to \(f\), then we may evaluate \(\widehat f_{N-J}\) as
\[\widehat f_{N-J} = H^J\boldsymbol a_{N-J} = \left(H^J\right)'H^J\boldsymbol x.\]

\paragraph{Finite sequences and edge effects}
In reality, \(\boldsymbol x\) is only finitely long, which can introduce issues at the boundaries. Suppose \(\boldsymbol x = (x_1,...,x_T)' = (a_{N,1},a_{N,2},\dots,a_{N,T})'\), and consider an analysis based on the Daubechies wavelet of order \(p\). If we wish to produce an approximation to \(\boldsymbol x\) at level \(N-J\), there will be exactly \(q=\lceil \frac{T+2(p-1)(2^J-1)}{2^J}\rceil\), scaling functions at level \(N-J\) whose support overlaps with the support of the dataset, \((1,2,...,T)\). In total, the support of these scaling functions is of length \(L=2(p-1)(2^{J}-1) + 2^Jq\). To deal with this we extend the dataset to be the appropriate length before applying a truncated version of the operators \(H\) and \(G\). In this work, we use a symmetric padding, where the dataset is reflected at \(t=1\) and \(t=T\). In particular, we place \(2(p-1)(2^{J}-1)\) extra terms at the beginning of the dataset, and \(L-T-2(p-1)(2^{J}-1) \) terms at the end of the dataset. 

Now to define the truncated versions of \(H\) and \(G\) to apply to this dataset. First, set \(s_0 = L\), \(s_1 = \lfloor \frac{L-1}{2}\rfloor+p \) then recursively define \(s_{j+1} = \lfloor \frac{s_{j}-1}{2}\rfloor+p\), \(j=2,3,..,J\). Then let \(H_j\) and \(G_j\) be the \(s_{j+1}\times s_j\) matrices
\begin{align*}
H_j=
\left[\begin{array}{ccccccccccccc}
h_0&h_1&h_2  & \hdots & h_{N-1} & 0& 0&\hdots& \hdots& \hdots& \hdots& 0\\
0&0&h_0& \hdots & h_{N-3} & h_{N-2}& h_{N-1}& 0&\hdots & \hdots& \hdots&0\\
\vdots&\vdots&\ddots&\ddots& \ddots&\ddots&\ddots& \ddots&\ddots&\ddots&\ddots&\ddots\\
0&0&\hdots&\hdots& \hdots&\hdots&\hdots&\hdots& \hdots& h_{N-1}&0&0\\
0&0&\hdots&\hdots& \hdots&\hdots&\hdots&\hdots& \hdots& h_{N-3}&h_{N-2}&h_{N-1}
\end{array}\right], 
\end{align*}
and 
\begin{align*}
G_j=
\left[\begin{array}{ccccccccccccc}
g_0&g_1&g_2  & \hdots & g_{N-1} & 0& 0&\hdots& \hdots& \hdots& \hdots& 0\\
0&0&g_0& \hdots & g_{N-3} & g_{N-2}& g_{N-1}& 0&\hdots & \hdots& \hdots&0\\
\vdots&\vdots&\ddots&\ddots& \ddots&\ddots&\ddots& \ddots&\ddots&\ddots&\ddots&\ddots\\
0&0&\hdots&\hdots& \hdots&\hdots&\hdots&\hdots& \hdots& g_{N-1}&0&0\\
0&0&\hdots&\hdots& \hdots&\hdots&\hdots&\hdots& \hdots& g_{N-3}&g_{N-2}&g_{N-1}
\end{array}\right].
\end{align*}

Thus we may compute coefficients \(\boldsymbol a_{N-J}\) and \(\boldsymbol b_{N-J}\) via
\[\boldsymbol a_{N-J} = H_{N-J} H_{N-J+1} \dots H_1\widetilde{\boldsymbol x},\]
and
\[\boldsymbol b_{N-J} = G_{N-J} H_{N-J+1} \dots H_1\widetilde{\boldsymbol x},\]
where \(\widetilde{\boldsymbol x}\) is the extended dataset. 

Similarly to before, the smoothed version of the signal is given by 
\[\widehat f_{N-J} = \left(H_{N-J} H_{N-J+1} \dots H_1\right)'H_{N-J} H_{N-J+1} \dots H_1 \widetilde{\boldsymbol x}.\]

\paragraph{An ordinary least squares regression view of wavelet filtering}
Wavelet filtering can also be viewed as an ordinary least squares regression problem. Consider the regression model
\[ \boldsymbol x = \left(H_{N-J} H_{N-J+1} \dots H_1\right)' \boldsymbol a + \sigma \varepsilon_t,\]
where \(\varepsilon_t\sim N (0,1)\). The least squares estimate of \( \boldsymbol a\) is 
\begin{align*}\widehat{\boldsymbol a} &= \left[\left(H_{N-J} H_{N-J+1} \dots H_1\right)\left(H_{N-J} H_{N-J+1} \dots H_1\right)'\right]^{-1}\left(H_{N-J} H_{N-J+1} \dots H_1\right)\boldsymbol x
\\&= \left[H_{N-J} H_{N-J+1} \dots H_1H_{1}'  \dots H_{N-J+1}'H_{N-J}'\right]^{-1}\left(H_{N-J} H_{N-J+1} \dots H_1\right)\boldsymbol x
\\&= H_{N-J} H_{N-J+1} \dots H_1 \boldsymbol x
\\&= \boldsymbol a_{N-J}, \end{align*}
due to the orthogonality of the columns of \(H_j'\).

The estimated line of best fit to the data is given by 
\[\left(H_{N-J} H_{N-J+1} \dots H_1\right)'\widehat{\boldsymbol a}=\left(H_{N-J} H_{N-J+1} \dots H_1\right)'\boldsymbol a_{N-J}=\widehat f_{N-J}.\]

Since the wavelet functions at resolution \(N-J\) and above are orthogonal to the scaling functions at resolution \(N-J\) then we do not need to include columns in the design matrix corresponding to wavelet functions in the regression model when all we wish to estimate is a smoothed trend. That is, the inclusion of wavelet components in the model would not affect estimates of \(\widehat{\boldsymbol a}\), and therefore not affect the estimate of the smooth trend.

\paragraph{A wavelet filtering inspired model for trends in a Bayesian setting}
In this paper, we consider a Bayesian setting to infer model parameters. Hence, we wish to translate the wavelet trend model described above into a Bayesian setting. With the view of wavelet filtering as a regression model, this becomes relatively simple. We include the columns of the design matrix \(W_j'=\left(H_{N-J} H_{N-J+1} \dots H_1\right)'\), in a design matrix for the trend component of our model. Specifically, we consider a model constructed using the Daubechies-8 wavelet, with \(J=8\) levels of smoothing.

\subsection{Cubic splines trend model}
A cubic spline is a piecewise cubic polynomial, which is continuous and has continuous first and second derivatives. Let \(\boldsymbol x = (x_0,\dots,x_T)\) be a dataset and \(\eta_1,\dots,\eta_k\) be \emph{knots}, where \(\eta_1<\eta_2<\dots< \eta_k\). The knots define the intervals over which cubic splines are piecewise cubic, \([\eta_1,\eta_2],[\eta_2,\eta_3],\dots,[\eta_{k-1},\eta_k]\). Within each of these intervals, cubic splines are infinitely differentiable. At the boundaries, however, we place restrictions on the basis functions so that the cubic splines have continuous first and second derivatives at the knots. We choose to use \emph{B-splines} to construct the basis functions which define the cubic splines. There are many equivalent bases that can be used for representing cubic splines, but B-splines are numerically stable, since they avoid calculating powers of large, or small, numbers. 

Before constructing our cubic B-spline basis functions, we first need to define an augmented set of knots, \(\tau_1,\tau_2,\dots,\tau_{k+2M}\). Set \(\tau_1=\tau_2=\dots=\tau_M<0\); set \(\tau_M<\tau_{M+1}=\eta_1, \tau_{M+2}=\eta_2,\dots,\tau_{M+k} = \eta_k<\tau_{M+k+1}\); and set \(T<\tau_{M+k+1} = \tau_{M+k+2} = \dots = \tau_{2M+k}\). B-splines are constructed recursively. Let \(B_{i,m}(x)\) be the \(i\)th basis function of an \(m\)-spline function, \(m\leq M\), with \[B_{i,1}(x) = \begin{cases} 1 & \mbox{ if } \tau_i\leq x < \tau_{i+1}, \\ 0 & \mbox{ otherwise,}\end{cases}\]
for \(i=1,\dots,k+2M-1\). Then \(B_{i,m}(x)\) can be constructed recursively via 
\[B_{i,m}(x) = \cfrac{x-\tau_i}{\tau_{i+m-1}-\tau_i}B_{i,m-1}(x) + \cfrac{\tau_{i+m}-x}{\tau_{i+m}-\tau_{i+1}}B_{i+1,m-1}(x),\]
for \(i = 1,\dots,k+2M-m\). We are interested in cubic splines, hence the basis functions we require are \(B_{i,4}(x)\) for \(i=1,\dots,k+4\). Note that we have some repeated knots, which would mean dividing by 0. To avoid this, we define \(B_{i,1}=0\) when \(\tau_{i}=\tau_{i+1}\). 

For a cubic spline with \(k\) knots, \(\eta_1,\dots,\eta_k\), there are \(k+2\) parameters to be estimated using a B-spline basis. The number and location of the knots need to be chosen, which is a model selection problem. To simplify matters, we use 13 equally spaced knots in this work, which roughly corresponds to 1 knot every 180 days; we also set \(\eta_1=-1\) and \(\eta_k=T+1\).

The observations in the dataset used here are taken at times \(0,1,\dots,T\), hence the design matrix looks like 
\begin{align*}
C' = \left[\begin{array}{cccc} 
B_{1,4}(0) & B_{2,4}(0) & \hdots & B_{k+4,4}(0)\\
B_{1,4}(1) & B_{2,4}(1) & \hdots & B_{k+4,4}(1)\\
\vdots & \vdots & \vdots & \vdots \\
B_{1,4}(T) & B_{2,4}(T) & \hdots & B_{k+4,4}(T)
\end{array}\right].
\end{align*}

\subsection{Weekly component trend model}
Another element of the trend component, is the short-term periodic component. This component captures a day-of-week effect observed in prices. The model we use is 
\begin{align*}g_t &= \beta_\text{Mon}\mathbb I (t \in \text{Mon}) + \beta_\text{Tue}\mathbb I (t \in \text{Tue}) + ...+ \beta_\text{Sun}\mathbb I (t \in \text{Sun}),\end{align*}
where \(\beta_\text{Mon},\beta_\text{Tue},...,\beta_\text{Sun},\) are parameters to be estimated relating to average prices on Monday, Tuesday, ..., Sunday, respectively. In a regression context, this model can be encoded as the columns of the following design matrix; 
\(M' = \boldsymbol 1\otimes I_7,%left[\begin{array}{ccccccc}1&0&0&0&0&0&0\\0&1&0&0&0&0&0\\0&0&1&0&0&0&0\\0&0&0&1&0&0&0\\0&0&0&0&1&0&0\\0&0&0&0&0&1&0\\0&0&0&0&0&0&1\\1&0&0&0&0&0&0\\\vdots&\vdots&\vdots&\vdots&\vdots&\vdots\end{array}\right].
\) 
where \(\boldsymbol 1\) is a vector of ones \(I_7\) is the \(7\times7\). That is, \(M'\) contains the identity matrix repeated as many times as required so that \(M'\) has \(T\) rows. 

\subsection{Discussion: putting it all together in a Bayesian setting}
Let \(Z\) be the design matrix of some trend model, and \(\boldsymbol z_k\) be the row of \(Z\) corresponding to the time point \(t_k\). Given a vector of parameters \(\boldsymbol \gamma\) for the trend model, the estimate of the trend at time \(t_k\) would therefore be \(\boldsymbol z_k \boldsymbol \gamma\). 

For the cubic spline model, the matrix \(Z\) is simple to construct given the work above; \(Z=[M'|C']\), that is, the matrices \(M'\) and \(C'\) concatenated. The wavelet-based model is slightly trickier, since we need to account for edge effects using padding. The design matrix for the wavelet trend component includes extra rows at the top, and the bottom, corresponding to padding. Specifically, since we use the Daubechies-8 wavelet and \(J=8\) levels of smoothing, there are \(\ell_1=3570\) rows at the top, and \(\ell_2=3802\) rows at the bottom, of the matrix \(W'\) which align with padding. So that we may concatenate the matrices \(M'\) and \(W'\), we pad the columns of \(M'\) via the same symmetric extension we use to pad the dataset. The full design matrix for the wavelet-based trend model is \(Z=[M^*|W']\), where \(M^*\) is the padded version of \(M'\). Only when estimating parameters corresponding to coefficients of scaling functions in the wavelet model do we use the extra rows in the design matrix corresponding to padding. In this setting, padding can be thought of as a kind of \emph{regularisation}, keeping the model from overfitting at the boundaries.

\subsection{Details of the MCMC Implementation} 
The Metropolis-Hasting (MH) algorithm is a very popular MCMC algorithm. For a target density \(p(\boldsymbol\theta | \boldsymbol x) \propto p(\boldsymbol x | \boldsymbol\theta')\pi(\boldsymbol\theta')\) the Metropolis-Hastings algorithm constructs a Markov chain that converges to the target as follows. 
\begin{enumerate} 
\item Propose that the chain move to a new position \(\boldsymbol\theta'\) which is drawn randomly from a proposal distribution centred at the current state of the chain \(\boldsymbol\theta^{(n)}\).
\item Calculate the ratio 
\[\lambda(\boldsymbol\theta',\boldsymbol\theta^{(n)}) = \frac{p(\boldsymbol x | \boldsymbol\theta')\pi(\boldsymbol\theta') R(\boldsymbol\theta^{(n)}|\boldsymbol\theta',s)}{p(\boldsymbol x | \boldsymbol\theta^{(n)})\pi(\boldsymbol\theta^{(n)}) R(\boldsymbol\theta'|\boldsymbol\theta^{(n)},s)},\]
where \(p(\boldsymbol x | \boldsymbol\theta)\) is the likelihood evaluated with parameter \(\boldsymbol\theta\), \(\pi(\boldsymbol\theta)\) is the prior, and \(R(\cdot| \boldsymbol\theta,s)\) is the proposal density centred at \(\boldsymbol\theta\).
\item Generate a random number \(u \sim U([0,1])\) and if \(\lambda(\boldsymbol\theta',\boldsymbol\theta^{(n)})<u\) set \(\boldsymbol\theta^{(n+1)} = \boldsymbol\theta'\); else, set \(\boldsymbol\theta^{(n+1)} = \boldsymbol\theta^{(n)}.\) 
\end{enumerate}

The choice of proposal distribution for the Metropolis-Hastings algorithm is theoretically arbitrary, up to some not very restrictive sufficient conditions (see \cite[Chapter 7]{robert2004} for sufficient conditions), in that the chain converges to the stationary distribution no matter what proposal distributions we use. However, if the chain is to reach stationarity in a reasonable amount of time, we must choose `good' proposal distributions. This is the justification for the block structure and adaptive steps in our MCMC implementation. %Also, notice that is a symmetric proposal distribution is used then the ration of the two proposal distributions is 1, also if a uniform prior is used then the ratio of the priors is also 1. The acceptance ratio \(\lambda\) is then just the ratio of the likelihoods. We use this fact in our implementation. 

The block Metropolis-Hastings algorithm can be seen as an extension of the Gibbs sampler: the sequential update structure of a Gibbs sampler is used but instead of sampling parameters directly from their conditional posterior, Metropolis-Hastings-type steps are used to update the chain. To implement a Gibbs sampler, the parameter vector is partitioned into blocks; \(\boldsymbol\theta = (\boldsymbol\theta_1,\boldsymbol\theta_2,...,\boldsymbol\theta_\ell)'\). The hidden regime sequence is also partitioned into blocks. Each element of the regime sequence, \(R_t\), is a block, \(\boldsymbol R = (R_1,...,R_T)'\). The Gibbs sampler iterates block-by-block sequentially updating the block by sampling from the conditional posterior distributions \(p(\theta_i | \theta_1,...,\theta_{i-1}, \theta_{i+1},...,\theta_\ell, \boldsymbol x , \boldsymbol R)\) for \(i=1,...,\ell,\) and \(p(R_t | \boldsymbol x , \boldsymbol \theta, R_1,...,R_{t-1}, R_{t+1},...,R_T)\) for \(t= 1,...,T\). Once each block is updated, the Gibbs sampler goes back to the first block are repeats the sequential update process. We refer to a single update of all the blocks as a \emph{sweep} of the algorithm. 

In some cases the Gibbs conditional distributions will be expensive to compute, or even intractable. A solutions it to propose updates to each block using a Metropolis-Hastings-type update rather than a Gibbs update. Hence the conditional posterior distributions do not need to be computed and performance of the algorithm may be improved. For the models considered here Metropolis-Hastings updates for the regime sequence are significantly faster than constructing and sampling from the conditional posteriors, \(p(R_t |  \boldsymbol x , \boldsymbol \theta, R_1,...,R_{t-1}, R_{t+1},...,R_T)\). In contrast, sampling from the conditional distributions \(p(p_{i1},...,p_{iN}|\boldsymbol x, \boldsymbol \theta, \boldsymbol R^{(n)} )=p(p_{i1},...,p_{iN}|\boldsymbol R^{(n)} )\) is relatively efficient. % In a block MH algorithm each block is updated sequentially using MH-style moves. That is, to get from \(\theta^{(n)}\) to \(\theta^{(n+1)}\) we update each block in turn, \(\theta^{(n)} = (\theta_1^{(n)},\theta_2^{(n)},...,\theta_l^{(n)}) \rightarrow (\theta_1^{(n+1)},\theta_2^{(n)},...,\theta_l^{(n)}) \rightarrow (\theta_1^{(n+1)},\theta_2^{(n+1)},...,\theta_l^{(n)}) \rightarrow ... \rightarrow (\theta_1^{(n+1)},\theta_2^{(n+1)},...,\theta_l^{(n+1)})=\theta^{(n+1)}  \) using MH steps.

For the models in this paper, the parameter vector \(\boldsymbol \theta\) contains the parameters for the trend component, \(\boldsymbol \gamma\), as well as appropriate subsets of the parameters \(\phi_i,\, \sigma_i\), \(i=1,2\), \(\mu_i,\, \sigma_i,\, q_i,\) \(i = 3,4,5\), and \(p_{ij}\) for \( i,j \in \mathcal S\) for the transition matrix of the regimes, depending on which regimes are included in the model. Each individual element in \(\boldsymbol \gamma\) and each individual parameter \(\phi_i,\, \sigma_i\), \(i=1,2\), \(\mu_i,\, \sigma_i,\, q_i,\) \(i = 3,4,5\) are blocks, and each row of the transition matrix, \(p_{i1},...,p_{iN}\), is also a block. Since
\(p_{iN} = 1-\sum\limits_{k \in \mathcal S} p_{ik},\)
we need to infer \(N(N-1)\) parameters for the transition matrix only. %To improve readability in the following presentation, we assume there are two regimes only, an AR(1) base regime and an i.i.d.~spike regime. The same structure of the algorithm applies for an arbitrary number of regimes. For two regimes the parameters are \(\boldsymbol\theta = (p_{11},p_{22},\alpha_1,\phi_1,\sigma_1,\mu_2,\sigma_2,q_2)\). 
%Let the vector \mbox{\(\boldsymbol R = (R_1,R_2,...,R_T)' \)} denote the regime sequence where the state space for each \(R_t\) is \(\mathcal S\). 
%Each sweep of the algorithm updates the state of the MCMC chain from \((\boldsymbol\theta^{(n-1)},\boldsymbol R^{(n-1)})\) to  \((\boldsymbol\theta^{(n)},\boldsymbol R^{(n)})\).

\paragraph{Proposal Distributions} 
At each sweep of the algorithm, the parameters \(p_{ij}\) are updated using a Gibbs-type update. That is, for each \(i\in\mathcal S\), we sample \(p_{i1},...,p_{iN}\) from \cite{henneke2011}
\begin{equation}\label{gibbspii}
p(p_{i1},...,p_{iN}| \boldsymbol R^{(n)} ) \sim Dirichlet(\eta_{i,1}+1,...,\eta_{i,N}+1),
\end{equation}
where \(\eta_{i,j} \) is the number of transitions from state \(i\) to state \(j\) in the current regime sequence \(\boldsymbol R^{(n)}\).

At each sweep of the algorithm, \(10\%\) of indices from the set \(I = \{1,2,...,T\}\) are randomly selected and for each sampled index \(t\) we update \(R_t\) as follows using a uniform distribution. Suppose the current regime sequence is \(\boldsymbol R^{(n)}\) and \(R_t^{(n)} = i\). Uniformly propose a move to any other state \(R_t' = j \neq i\), and accept or reject this move with the appropriate probability as given by the Metropolis-Hastings algorithm. By only choosing \(10\%\) of the elements in hidden sequence to update at each sweep of the algorithm, rather than update all \(T\) elements, we greatly decrease the running time of the algorithm, and yet this does not significantly affect the mixing of the MCMC chain.

At each sweep of the algorithm, given the current state of the MCMC chain, \(\boldsymbol \theta^{(n)}\), the \(k\)th block of the parameter vector is updated via proposing a move from a normal distribution with mean \(\theta_k^{(n)}\) and variance \(s_k^2\). The move are accepted or rejected according the a Metropolis-Hasting-type acceptance probability. The proposal variances \(s_i\) are found using an adaptive algorithm which we discuss later. 

\paragraph{Adaptive proposal distributions}
Roberts and Rosenthal \cite{roberts2001, roberts2009} provide an adaptive algorithm to find effective variance parameters, \(s_k^2\), for an MCMC algorithm similar to ours. In \cite{roberts2001} Roberts and Rosenthal prove, for an idealised version of our block Metropolis-Hasting algorithm, an optimal acceptance rate is 0.44. In \cite{roberts2009} they provide an example of an adaptive scheme, which automatically adjusts the parameters \(s_k^2\) in the algorithm to asymptotically reach the optimal acceptance rate while maintaining the necessary ergodicity and convergence properties for the same idealised problem. The problem they considered has a posterior distribution that is a multivariate normal, whereas for our problems, it is unlikely that the posterior distributions are normal, and in addition we also have to sample the hidden regime sequence. Nonetheless, this adaptive scheme works well for our purposes.

Our implementation of their adaptive scheme is as follows. For each parameter we initialise the standard deviation of the proposal to \(s_k=1\) and begin our block Metropolis-Hastings algorithm. After the \(n\)th batch of 50 iterations of the block Metropolis-Hastings algorithm, we update \(s_k\) by multiplying by \(\exp(\delta(n))\) if the acceptance rate is above 0.44, or by \(\exp(-\delta(n))\) if the acceptance rate is less than 0.44. Following the ideas in \cite{roberts2009}, we define \[\delta(n) = \min\left(\frac{2}{\sqrt n},\frac{10}{n},\frac{10000}{n^2}\right).\] Note that to satisfy the conditions for convergence of this algorithm outlined in \cite{roberts2009}, we also need to specify a bound \(M<\infty\) and restrict \(\log(s_k)\) to \([-M,M]\). % (the condition \(\log(s_i)\in [-M,M]\) may not even ensure converge for this adaptive algorithm since in our problem we also sample \(\boldsymbol R\)). 

\end{document}